 \documentclass[authoryear,preprint,review,12pt]{elsarticle}

\newcommand{\beq}{\begin{equation}}
\newcommand{\eeq}{\end{equation}}
\newcommand{\no}{\noindent}

\usepackage{amssymb}

\journal{Earth-Science Reviews}

\begin{document}

\begin{frontmatter}



\title{Short-Term Orbital Forcing: A Quasi-Review and a Reappraisal  of  Realistic Boundary Conditions for Climate Modeling}


\author[rvt]{Rodolfo G. Cionco\corref{cor1}}
\ead{gcionco@frsn.utn.edu.ar}
\address[rvt]{Comisi\'on de Investigaciones Cient\'ificas de la Provincia de 
Buenos Aires --  Universidad Tecnol\'ogica Nacional, Col\'on 332, 
San Nicol\'as (2900), Bs.As., Argentina}
\cortext[cor1]{Corresponding author}
\author[focal]{Willie W.-H Soon}
\ead{wsoon@cfa.harvard.edu}
\address[focal]{Harvard-Smithsonian Center for Astrophysics, Cambridge, Massachusetts, 02138, USA}

\begin{abstract}
The aim of this paper is to provide geoscientists with the most accurate set of the Earth's astro-climatic parameters and daily 
insolation quantities, able to describe the Short-Term Orbital Forcing (STOF) as represented by the ever-changing incoming solar radiation. We 
provide an updated review and a pragmatic tool/database using the latest astronomical models and orbital ephemeris, for the entire Holocene 
and 1 kyr into the future. Our results are compared with the most important database produced for studying long-term orbital forcing 
showing no systematic discrepancies over the full thirteen thousand years period studied.   
 Our detailed analysis of the periods present in STOF, as perturbed by Solar System bodies, yields  a very rich dynamical modulation 
 on annual-to-decadal timescales when compared to previous results.
 In addition, we addressed, for the first time, the error committed considering daily insolation as a continuous function of  
orbital longitudes with respect to the nominal values,  i.e.,    
calculating the corresponding daily insolation with orbital longitudes tabulated {\it at noon}.
We found important relative differences up to $\pm$ 5\%, which correspond to errors of 2.5 W m$^{-2}$ 
in the daily mean  insolation, for exactly the same 
calendar day and set of astro-climatic parameters. 
This previously unrecognized error could have a significant impact in both the initial and boundary conditions for any climate modeling 
experiment.

\end{abstract}


\end{frontmatter}


\section{Boundary conditions for climate system: A brief historical review of orbital modulations of incoming solar radiation}
\label{Sec1}

The historical development for a scientific understanding of weather and climate change, particularly the astronomical theory of climate 
change is a long one \citep{Neumann1985}. For example, the Roman writer on agriculture, Lucius Columella (ca. 4 to 70 AD) began in his Book 
I De Re Rustica [On Agriculture]:

{\it
``For I have found that many authorities now worthy of remembrance were convinced that with the long wasting of the ages, weather and climate
undergo a change; and that among them the most learned professional astronomer, Hipparchus, has put it on record that the time will come when the
poles will change position, a statement to which Saserna, [or Sasernas {\rm because J. Neumann noted that this is most likely the writing from a 
father and son team}] no mean authority on husbandry, seems to have given credence. For in that book on agriculture which he has left behind he 
concludes that the position of the heavens had changed from this evidence: that regions which formerly, because of the unremitting severity of 
winter, could not safeguard any shoot of the vine or the olive planted in them, now that the earlier coldness has abated and the weather is 
becoming more clement, produce olive harvests and the vintages of Bacchus in the greatest abundance. But whether this theory be true or false, we 
must leave it
to the writings on astronomy.''}\footnote{The full text of this book is available at {\tt http://penelope.uchicago.edu/Thayer/E/Roman/Texts/
Columella/de\_Re\_Rustica/1*.html}.} 

Fast-forward to modern scientific age, the well-known calls for ``weather forecasting as a problem in physics" and ``climate as a problem of
physics" \citep[c.f.,][]{Monin2000} as well as the recent quest for ``a theory of climate'' \citep{Essex2011} clearly exemplify and highlight 
the struggle of the science of meteorology and climatology for the past century till today. We may echo the most pristine quest set out by one M.
Milankovi\'c, among other accomplishments also conferred as the father of climate modeling by \cite{Berger2012} upon a comprehensive review 
of the subject, that:

{\it
`` \ldots such a theory would enable us to go beyond the range of direct observations, not only in space, but also in time\dots It would allow 
reconstruction of the Earth's climate, and also its predictions, as well as give us the first reliable data about the climate conditions on other 
planets.''}\footnote{From p. 6 of \cite{Petrovic2009} where the first sentence of the quote has original words taken directly from Milankovi\'c's 
autobiography. We wish to note that the Wikipedia entry for Milutin Milankovi\'c 
{\tt https://es.wikipedia.org/wiki/Milutin\_Milankovi\%C4\%87}
has incorrectly implied the quote to be from a 1913 paper entitled ``Distribution of the Sun radiation on the Earth's surface" (in Serbian).} 

Such facts and reality alone would demand an ever increasing scrutiny for a better precision and accuracy in defining and specifying what has been
well accepted to be the so-called external forcing boundary conditions for Earth's complex and coupled climate system which involves not only the
atmosphere, land and ocean per se but must also necessarily includes a host of other fast, slow and intermediate scales processes involving
both the intrinsic magnetic variations and modulation of the Sun's radiation outputs and the distinct effects from gravitational interactions of
all bodies in the Solar System (SS) in affecting the geometrical variations of the Moon-Earth's orbital elements. This basic requirement for a 
correct
boundary condition for climate modeling can be sharpened further if climate variations are considered as a characteristic to be deduced from
rather than artificially imposed on a realistic climate model \citep[e.g., ][]{Nicolis1995,Lions1997,Goody2007}.

It is clear from a careful literature review (see detailed discussion below for a brief review of those longer term orbital changes) that the 
limited availability of STOF parameters has thus far prevented from a more direct prescription of this
particular boundary condition in any meteorological and climatic simulations and studies of the past, present and future. To the best of our
knowledge, a rare exception is contained in the preliminary study by \cite{Bertrand2002}. A close inspection from the United Nations Fifth
Assessment Report \citep[see][p. 1051]{Collins2013} tells us that STOF is indirectly assumed to be unimportant and play no climatic role. The
primary argument and assumption in neglecting changes in orbital forcing for climatic changes over the last few thousand years, last century or
even last decade seemed to be from the claim that {\it globally-averaged} radiative forcing is small or negligible. Other reasoning, involved in 
climate reanalysis projects and products \citep[see e.g., ][]{Simmons2014,Hersbach2015}, seemed to be that such an effect involved in the
modulations of the seasonal forcing and seasonality itself has been indirectly included in other boundary conditions like through the prescription
of sea surface temperatures or atmospheric ozone concentration or through assimilation of other atmospheric metrics in a 100-year atmospheric
reanalysis simulation. 

More than two decades ago, \cite[][p. 525]{Laskar1993} echoed a reminder that:

{\it 
``The mean annual insolation depends thus only on the eccentricity of the Earth, and its variation over 1 Myr are \dots [{\rm very small, about 
0.6  W m $^{-2}$ as shown in Figure 4 of Laskar et al. 1993}] \dots In fact, this is not the main paleoclimate quantity, and it was recognized by 
[{\rm Milankovi\'c}]
\dots that the summer insolation at high latitudes had a larger influence on the climate of the past. If the insolation in summer is not high 
enough the ice does not melt, and the ice caps can extend. This is why it is also important to compute the daily insolation at a given point on 
the Earth.''}

We agree and only wish to call for a more direct accounting for STOF, as a true boundary condition, in all climatic simulations in that the
effects from local and regional perspective are clearly not negligible nor unimportant \citep[see e.g., section 3.7 in][]{MillerSchmidt2014} in 
terms of seasonal dynamical evolution of the coupled air-sea-land system. The importance of accurately specifying the incoming solar radiation on 
local and regional
geographic scales should also be studied and assessed in contexts with the currently unresolved difficulties in closing the budget of Earth's 
energy imbalance 
\citep{Trenberth2016} as well as in both interpreting and emulating local and regional surface temperatures using the state-of-the-art general 
circulation models \citep{JiWu2014,LinHuybers2016}.  Furthermore, the persistent modulation of the orbital forcing, via the modulation
of the seasonal irradiation amplitudes, ranges and their geographical distributions, all across the globe may yield a deeper insight to the basic
problems of climate science on timescales of decade to century.

We shall limit our exploration and calculation to the Holocene and 1 kyr into the future, for which the longest available accurate ephemeris
including this period is the Jet Propulsion Laboratory's Development Ephemeris 431 (JPL's DE431) outputs \citep{Folkner2014}. 
Another simplification we adopt for this paper is the postponement of the accounting of the
intrinsic variations of the Sun's irradiance outputs as caused by the variable nature of solar magnetism on a host of timescales but a recent
review on the choices of Total Solar Irradiance (TSI) records to use for climatic study and modeling has been reported in \cite{Soon2015}.

We will not further speculate on the importance of getting the correct boundary conditions for any atmospheric, meteorological and climatic 
simulations, but will limit ourselves to a brief note on some recent works as a source of motivation. \cite{Cronin2014} has recently highlighted 
the key role of getting the solar zenith angle correctly represented in global climate models where errors in assuming daytime-average zenith 
angle or spatially constant insolation may lead to underestimating or overestimating of global energy budget on the order of 10 W m$^{-2}$ or 
more, respectively.
\cite{Hogan2016} documented the situations when the infrequent calls to radiation module in a climate model calculation may lead to  
errors in stratospheric temperatures of 3 to 5 K. \cite{He2015} spelled out the detailed dependence of the amount of solar radiation on 
the length of day and latitude which in turn for defining the role of Tibetan and Iranian Plateau in constraining the northerly intrusion into the
Indian monsoon area during boreal summers.The multi-scale roles of solar radiation on the diurnal
 cycles and related coupled ocean-atmosphere dynamics of the warm surface ocean including tropical Indian Ocean and the western
 equatorial Pacific have been examined by \cite{Shinoda2005} and \cite{LiShinoda2013}.
 \cite{Hudson2016} highlighted the key role by solar irradiation impacts for the reflection heights in the ionosphere
 that are changing by 15 km, 12 km and 1 km, respectively, by diurnal, seasonal and 11-yr cycles of solar irradiation.  
\cite{Jajcay2016} shows and discusses the evidence on how the small-amplitude 7-8 yr 
oscillations in the European surface air temperature records can produce the so-called cross-scale modulation in the temperature at a higher
frequency range and regime, in their case the annual cycles. It is far from clear that the multi-scale STOF reality presented in our paper would
have absolutely no role for this case study by \cite{Jajcay2016}.
Finally, \cite{Joussaume1997} and \cite{Chen2011} have cautioned earlier on concerning
 the differences in calendar assumptions (i.e., differences in the Sun's positions with respect to calendar days) that can lead
to a rather significant impact in climate modeling through phase shifts on the prescribed solar irradiance values. This is why, as emphasized in 
this work, an accurate and reliable way of tracking time in the orbital calculation is important.    

\section{A quasi-review and overview of several important previous works}
\label{Sec2}

From a paleo-climatic perspective, the most widely applied 
studies 
related to changes in the incoming solar radiation (the so called insolation)
due to variations in both Earth's orbit and Earth's celestial pole, have been considered in detail by, e.g., \cite{Berger1978} and
references therein, \cite{BergerLoutre1991}, \cite{Laskar2004,Laskar2011} for a very long time-span (i.e., geologic eras). Such long-term
forcing, varying on tens to hundreds thousands of years on incoming solar radiation, is often called the 
{\it Milankovi\'c orbital forcing}. These works
are effort based on celestial mechanics publications of \cite{Milankovic1941}, \cite{Sharaf1967}, \cite{Bretagnon1974}, 
\cite{Berger1978}, \cite{Laskar1988}, and \cite{Laskar2004} among others. Both \cite{Imbrie1982} and \cite{Berger2012} proffered a comprehensive 
review on the history of the {\it orbital solutions} which now
have improved techniques and methodologies starting with the classical orbital problem of the evolution of the gravitational solar $N$-body 
system, which have their roots in the works of Laplace and Lagrange \cite[see e.g.,][]{Laskar2013}.

These solutions (as it is usually referred to in astronomical literature for the process in obtaining orbital elements and Earth's orientation 
parameters from a semi-analytic model of the SS evolution) are intended for several million years. Such calculations are based on the numerical 
integration of 
the equations which describes the secular (i.e., long-term and long-period) dynamics of the SS, where short-term 
periodicities, arising from the evolution of the orbital longitudes itself, were simply not taken into account in order to reduce the complexity 
of the orbital integrations.
Specially, the works by \cite{BergerLoutre1991} 
cover the field for $-$5 Myr (i.e., before present); \cite{Laskar2004} integrate a secular-realistic model of the SS, given accurate solutions 
for $-$50 Myr and +20 Myr from present. \cite{Laskar2011} also using very accurate initial conditions and very short time step (less than 
one year), have recently extended the previous solution to $-$250 Myr, i.e., including all the Cenozoic era.\footnote{In addition to what was 
already known about the chaotic evolution 
of orbital motions of terrestrial planets in the SS with a characteristic Lyapunov time of 5 Myr (i.e., largely limited by the lack of 
knowledge on the precise value of solar oblateness, $J_2$), it is worth noting that \cite{Laskar2011b} recently proposed that ``it will never be 
possible" to track the precise evolution of the Earth's eccentricity beyond 60 Myr owing to the strong chaos caused by close encounters or even a 
chance collision between Ceres and Vesta.} 
These solutions coming from works by Laskar and colleagues although permitting the {\it insolation quantities} at, e.g., annual timescale, but 
all those impressive results did not account for the STOF variations.

The key issue tackled in our present work is the problem of the short-term variations of insolation quantities; i.e., the high-frequency orbital 
forcing, and its variations at decadal and multidecadal scales. This problem was first addressed by
 \cite{Borisenkov1983} and \cite{Borisenkov1985}. They determined short timescale variations between $\sim$ 2-20 yr
which produce variations of the same order as the long-term orbital forcing in a few hundred years. They found that the greatest short-term effect
of orbital forcing occurs in high latitudes at midsummer (July); with the lunar orbital retro-grading period of 18.6 yr (i.e., the main nutation
term) as the most evident driver \citep[see e.g., the recent evidence found for this modulation of sea level at the Eastern North Sea and Central 
Baltic Sea regions by][]{Hansen2015}. This effect is found to be of the same order of amplitude as the intrinsic irradiance variations due to 
solar magnetic cycle variations. \cite{Loutre1992} performed the most complete analysis of short-term orbital forcing to date. They analyzed 
the Earth's parameters involved in orbital forcing, which are the so-called 
{\it astro-climatic elements} or parameters, for the last 6 kyr at an annual basis (i.e., getting one value for each calendar year). Their 
astronomical solution is based on the French VSOP planetary theory \citep{Bretagnon1982}. This semi-analytic theory, was developed in series up 
to third order of the masses of all the planets considered plus an improvement for the descriptions of the giant planets up to the sixth order. 
The lunar perturbation in the Earth-Moon barycenter is also included; but the Earth and the Moon are not considered as separated bodies. The 
basic integration constants used were adjusted from former JPL DE200 ephemeris \citep{Standish1982}. Therefore, this solution allowed 
considerably shorter periodicities in planetary perturbations than former astronomical solutions.
 \cite{Loutre1992} confirmed the earlier results of Borisenkov and collaborators, showing that the most important spectral peaks occur at 
decadal timescales (say, periods less than 40 yr),  but also specified longer periodicities, some of them at multicentennial to nearly millennial 
scales. They found that, for
longer periodicities the precessional modulation produced the most important signals for low to middle latitudes, with the obliquity signals
stronger at solstices for polar latitudes. Such an observation is clearly consistent with what is mostly known for the Milankovi\'c orbital
timescales \citep[see e.g.,][]{Imbrie1982}.

It would be premature for us to spell out and prove the direct relevance of all such STOF periodicities for any ranges of climatic measures
and metrics while the state of understanding and discussion in the literature remains still largely unsettled. This is mainly because there has 
been little, with the few exceptions reviewed here, study for STOF in terms 
of any actual manifestations in the real external boundary conditions 
that can be subjected to any systematic evaluations. 

Although Milankovi\'c forcing data are available \citep{BergerLoutre1991,Laskar2004, Laskar2011}, e.g., from the National Centers for 
Environmental Information of NOAA/NCEI [formerly known as NOAA/NCDC] \\
({\tt http://www.ncdc.noaa.gov/data-access/paleoclimatology-data/data \\
sets/climate-forcing}), \\ 
detailed STOF data and the corresponding insolation quantities are still largely missing from any scientific archives and literature. Secular
(smooth) values of insolation quantities can be obtained at centennial resolution using orbital solutions by Laskar and colleagues, from this web
page (but, as was mentioned, only strictly for Milankovi\'c's scales of oscillations in tens of millennia). To-date, there is no direct data made 
available publicly, e.g., from \cite{Loutre1992} solution, describing STOF and insolation quantities. Hence, we have decided to provide 
geoscientists with the most complete and accurate descriptions of the STOF astro-climatic parameters using DE431 ephemeris, the longest high 
precision ephemeris publicly available. 
High precision ephemerides have the most accurate description of short-term variations of Earth's orbit. 
Another important improvement (i.e., with respect to Loutre et al. 1992) worth
applying now is the use of the latest precession-nutation formulas to get the astro-climatic parameters expressed in a reference system of 
{\it the date} (i.e., with respect to the moving equinox). Although a large effect may not be expected in our studied time-span of the Holocene, 
 it is important to use these latest formulations if an accurate accounting of the 
 dynamics of orbital forcing is desired. 
The classical polynomial precession formulas \citep{Lieske1977} are only 
valid for a few centuries from J2000.0 \citep{Vondrak2011} and non-rigid Earth nutation series are now available \citep{Mathews2002} 
for our inclusion. 

From Milankovi\'c works to present, the incoming solar radiation has been reckoned in different ways (i.e., the so called insolation 
quantities: instantaneous, daily, mean mid-month, monthly mean, seasonal, etc.). We think that any arbitrary uses of insolation or derived 
quantities related to it, could be confusing to expert modelers and non-expert readers alike. The basic insolation definitions and calculations 
over arbitrary period of times are periodically 
reviewed and explained in terms of the most basic geometry or refined mathematical tools \citep[e.g., ][]{Berger1978,Laskar1993,Berger2010}. In 
this work we also feel that it is necessary to start reviewing the basic definition of insolation. They are based
on daily irradiation over a whole Earth's variable rotation day \citep{Berger1978,Berger2010,Borisenkov1985,Loutre1992} which is
the basis or foundation of all longer time-span calculations. 

Daily irradiation is the fundamental parameter of all insolation calculations and its formulation is derived considering the constancy; over a
full day, of astro-climatic elements and also the solar orbital longitude (therefore, the solar declination; see Section \ref{Sec3}), 
assuming that the error committed considering this constancy of orbital longitude is also negligible \citep{Berger2010}. 
Nevertheless, for practical purposes, daily insolation formulas are usually applied, considering the metric as a {\it continuous function of 
solar longitude}; i.e., as a continuous function of time, setting a desired longitude value, which, in fact, could be anything within a 
particular day. 
Such a procedure in adopting the daily insolation formulas as a continuous function of solar longitude, is an habitual practice in 
paleoclimates studies, where insolation variations at a particular orbital longitude (e.g., solstices, equinoxes, etc.) is followed along the 
ages. 
The error with respect to the nominal (i.e., a constant solar longitude) daily insolation formulas has not been addressed in the literature, 
although one can assume a level of accuracy of 0.01 W m$^{-2}$ 
in the mean daily insolation, 
following the claim in \cite{Berger2010}. 
In this work, we have tabulated accurate daily insolation quantities 
at periods of one day (one Julian day of 86400 s); so we can evaluate the difference or ``error'' in evaluating the same insolation quantities, 
adopting both approaches, calculating daily insolation formulas as a continuous function of orbital longitude and using tabulated values at noon 
of the corresponding day. 

The main aim of this work is to describe and calculate the short-term dynamics of the 
Earth's astro-climatic elements, with the highest precision available and the basic
daily insolation quantities suitable for climatic modeling, for the whole Holocene up to 1 kyr in the future. We present our 
 analyses adopting the solar true orbital longitude coordinate in order 
to avoid ``calendar-problems" related to ``fixed-day" calendar used in most paleoclimate studies 
\citep{Joussaume1997,Chen2011,Steel2013}. Nevertheless, we also express our data in Julian days; which is a standard time-reckoning in astronomy, 
this also provides a link to civil calendar of the past and future.  
Although STOF variations produce departures of the secular, long-term values less than 
$\pm$ 0.15 W m$^{-2}$, we found that calculating daily
insolation quantities with tabulated values for a certain day or using it as a continuous function of orbital longitude, can produces important
differences, up 
to $\pm$ 2.5 W m$^{-2}$ ($\pm$5\%) in the mean daily insolation, for the same set of astro-climatic parameters. This holds even for the present 
day. Again, 
along our theme of climatic variations should be a {\it deduced} characteristics of the Earth's coupled non-linear dynamical system, we consider 
that taking additional cares in accounting for this aspect of STOF boundary conditions in any climate simulation or attribution studies are
both necessary and important.

\section{Solar irradiance intercepted (or received) by the Earth and coordinate systems for studying STOF variations}
\label{Sec3} 

Let us define the instantaneous solar irradiance received at the top of the atmosphere above a geographic location on Earth, at
 a certain moment of a given date. 
The amount of solar energy flux, 
$Q_t$, on a horizontal surface per unit of time, depends on the received Total 
Solar Irradiance, TSI,  (i.e., the total energy received from the Sun) which, as a flux, depends on the inverse-squared-law to the source, for us 
the Sun-Earth distance $r$ (reckoned in astronomical units, au),  and on the inclination of sun beams as follows:

\beq
\label{Eq1}
\frac{ {\rm d}Q_t}{{\rm d}t} = I = {\rm TSI}(r) \, \cos z; \ \ 
   {\rm TSI}(r) = {\rm TSI_0} \,  \left ( \frac{1 {\rm au}}{r}\right )^2,  
\eeq

\no $I$ is the instantaneous received solar irradiance, the intensity of the vertical component of sunshine, or the incoming solar radiation
(insolation); in MKS system the insolation is reckoned in J/(s m$^{-2}$) or W m$^{-2}$. TSI$_0$ is the ``solar constant'', the solar irradiance
received at a fiducial distance by a surface directly oriented to the Sun. The angle subtended by the Sun from the zenith, i.e., the zenithal 
distance, $z$, can be expressed as functions of the latitude 
of the observer and of the declination of the Sun, $\delta$, as shown in Fig. 1. Here a local fixed geocentric equatorial system (X,Y, Z) is
defined, X directed towards the local meridian; $-$Z towards the hemispheric pole (south, here); therefore the direction of the Sun, 
$\breve{n}_{\odot}$, and the direction of the observer's zenith, $\breve{n}_z$, can be written as:

\beq
\label{Eq2}
\breve{n}_{\odot} = (\cos \delta  \, \cos H, \cos \delta  \, \sin H, \sin \delta); \ \ \breve{n}_z = ( \cos \phi , 0, \sin \phi),
\eeq

\no where $H$ is the hour angle (reckoned from the local meridian, following the sense of the Sun's apparent motion). To avoid confusion with sign
conventions, we assume that $\delta$ and $\phi$ are the absolute value of the celestial and terrestrial angles to depict the problem and to
consider in our formulation; in subsequent applications we must put the corresponding signs 
(in fact in Fig. 1, $\delta$ and $\phi$ need to be 
negative numbers to fulfill the right-hand rule of the coordinate axes). Besides, we neglect the diurnal parallax effects, so the
geocentric angles are equal to topocentric ones (this assumption leads to a correction at a maximum of 9 arcsec). It is thus straightforward that:

\beq
\label{cosz}
\cos z = \breve{n}_z \cdot \breve{n}_{\odot}= 
   \sin \phi \sin\delta + \cos \phi \cos \delta \cos H, 
\eeq

\no then we can write the flux of the solar energy projected on a horizontal plane, per units of time as:  

\beq
\label{dqdt}
\frac{{\rm d}Q_t}{{\rm d}t} = I = {\rm TSI}_0 \  \left ( \frac{1 {\rm au}}{r}  \right)^2 \   
(\sin \phi \sin\delta + \cos \phi \cos \delta \cos H  ).
\eeq

As a note, the ``solar constant" for us is the index received at 1 au 
(a fix distance to the Sun), 
e.g., 1364.5 $\pm$
1.38 W m$^{-2}$  \citep{Mekaoui2010}; 
but an estimation based on the mean Sun-Earth distance: $<r> = a \sqrt{(1-e^2)}$, where $a$ is the Earth's orbit semi-major axis and $e$ the 
Earth orbital eccentricity; 
produces a different factor of $(1-e^2)^{0.25}$ in Eq. (\ref{dqdt}).  The importance of getting the absolute value 
of TSI$_0$ index correctly for climatology has been recently discussed and explored in \cite{Soon2014}. For all the calculations in this paper, 
we adopted TSI$_{0} = 1366$ W m$^{-2}$ mainly to provide an apple-to-apple comparison with the numerical results shown in \cite{Loutre1992} 
but of course TSI$_0$ is left as an adjustable free parameter in our computer programs.

It is evident that for a fixed latitude, the insolation $I$ depends strongly on seasonal changes which 
are driven by annual solar declination variations, with the Earth-Sun distance virtually constant at intra-annual timescale, with respect to $
\delta$ variations.  
 This can be clearly seen by relating $r$ and $\delta$ with the  
Earth's astro-climatic parameters \citep{BergerLoutre1991,Loutre1992};   
the above mentioned eccentricity, $e$; the longitude of the perihelion, $\varpi$; and the obliquity of the ecliptic (Earth's obliquity), 
$\epsilon$:

\beq
\label{rdelta}
r = \frac{a(1-e^2)}{1 + e 
\cos(\lambda-\varpi)} ;  \quad \sin \delta = 
\sin(\lambda + 180^{\circ})  \sin {\epsilon}
\eeq

\no where $\lambda$ is the true longitude of the Earth on its orbit, a polar angle of position of the Earth reckoned from the equinox of 
reference ($\gamma_0$)  which sets the reference system on which these elements were 
originally depicted and calculated (see Fig. 2). 
In what follows, we consider the Earth's semi-major axis to be a non-constant parameter, because it has short-term variations (but not secularly 
growing, at least, under the considered Holocene interval of twelve to thirteen thousand years), hence this parameter will produce a slight 
variation on the 
solar radiation received (of course, if we assume solar constant evaluated at $r=a$; then $a$ disappears in the formulation). 

In paleoclimate studies, it is usual to set the Sun's longitude, which we have already emphasized, and is defined as $\lambda_{\odot}=\lambda 
+ 180$, and not the Earth's longitude, because the apparent geocentric solar movement is described. 
Hence, in order to identify the correct time of the year when solstices and equinoxes occur, i.e., when the Sun passes through them in its 
geocentric orbit, we have indicated in Fig. 2, the month of the year  (for the present calendar time) and the corresponding position of the Sun 
in its reflex orbit.
Here, the vernal equinox (origin of longitudes) corresponds to $\lambda_{\odot}= 0 $ deg. (March), the June solstice is at $\lambda_{\odot}= 90 $ 
deg. (i.e., the Earth is near its aphelion), the equinox of September is at $\lambda_{\odot}= 180$ deg., and the solstice of December $\lambda_
{\odot}= 270 $ deg. (i.e., when the Earth is near its perihelion).  

At this point, it is important to remember 
that day by day, the origin of orbital longitudes, the $\gamma$ point, makes a complex movement involving components of longer period 
(precession) and shorter period (nutation). 
Such movements are produced by the combined planetary and luni-solar torques which, in turn, affect the
reckonings of longitude and also the Earth's obliquity (see Fig. 2 and the next section); i.e., the Earth's
pole positions on the celestial sphere.

Eq. (\ref{rdelta}) clearly 
shows that the movement of the origin of longitudes has a dramatic effect on the apparent positions of the Sun in the sky with respect to the 
seasons. 
Because of this, we need to keep the synchronization of the orbital longitudes with the seasons. 
Therefore, the origin of longitudes needs to be measured from the ``true --i.e, real and moving-- equinox of the date" ($\gamma$); hence if we 
call $\lambda_t$ the true longitude and $\varpi_t$ the longitude of the perihelion of the Earth measured from the {\it true equinox  of the date}, 
we must replace them in Eq. (\ref{rdelta}); then the trigonometric part of both equations become $\cos(\lambda_t-\varpi_t)$, and,
$\sin(\lambda_{\odot t})$, with $\lambda_{\odot t}$ as the true solar longitude {\it of the date}. 
 For the reference to calendar date (Gregorian, leap year mechanics),  time is kept following the precession,
 i.e., it is set, as far it is possible, to tropical year, then the
seasons are set with the months of the year at least for several thousand years before and after present. Nevertheless, inconmensurabilities
between the involved periodicities and proper irregularities of the changing Earth orbital elements, produces that fixed values of $\lambda_t$ 
does not occurs exactly at the same calendar date. In other words, the duration of the season varies through time. In paleoclimate, a 
{\it conventional} calendar is used, it is the present Gregorian calendar, with the present month versus Earth's longitude relationship. 
Nevertheless, 
for the above-mentioned reasons (fixed values of  longitude do not occur exactly at the same date), important differences in insolation 
estimates can occur when calendar date is used as temporal reference, even with the conventional calendar 
\citep[see below]{Joussaume1997,Steel2013}.

In what follows, we explain how to obtain high precision astro-climate elements from DE431 for the studied time span. Because we used a standard
ephemeris, we obtain the basic data at a certain fix time step, based on a fixed Earth's day of 86400 s, the Julian day. This is the standard way 
by which the SS bodies are positioned at a certain moment. It is interesting to note that this will also provide an exact link to the 
civil (Gregorian -- Julian) calendar. 

\section{Some astrometry}
\label{Sec4}

With the advent of fast computers, it is possible to resolve, by direct numerical integration, the set of differential equations which describe 
very accurate models of the planetary problem in the SS using high precision initial conditions, for a very long time span.
Moreover, the integrated orbits can be fitted to observations which improves the final precision of the solution. 
 In this 
way, we can get for instance, the heliocentric position, $\mathbf{r}$, and velocity, $\mathbf{v}$, of the Earth as a function of time; this 
process is referred to as the {\it ephemeris calculation}. 
DE431 ephemeris models the Earth's center of mass movement in the SS and some  
parameters of Earth's orientation in the inertial space. The position and velocity of the Earth are obtained by adopting the 
relativistic-barycentric reference system, related to the barycenter of the SS. 
Later, for a specific moment, through an {\it orbit calculation}, we can replace this set of six components of $
\mathbf{r}$ and $\mathbf{v}$ with six ``constants" of motion, the Keplerian elements, which have the convenient property for describing the 
orbital ellipses of the Earth around the Sun. 
Even more nicely, we can describe the apparent geocentric orbit of the Sun around the Earth with such a set of
elements. Of course, through time, these ``constants" vary because there are more than two bodies (some of them significantly non-spherical 
bodies) orbiting and interacting. Then, we need to follow through time the Earth's orbital elements as well as the orientation of the Earth 
celestial pole in space. 
These variations are the geometrical causes of insolation variations.

As mentioned, the longest high precision ephemeris publicly available is the JPL's DE431 integration (Folkner et al., 2014). It is designed to 
cover from  13,200 years before the reference epoch J2000.0 to 17,191 years after. 
 And DE431 is in good agreement with all the main ephemerides, 
i.e., the French INPOP \citep{Fienga2008} and  the Russian EPM ephemeris \citep{Pitjeva2014}. 
Unlike semi-analytical planetary theories such as VSOP \citep{Simonetal2013}, 
numerically integrated ephemerides naturally include {\it all} the terms of the disturbing functions; and integrates separately the Earth, 
from the Earth-Moon barycenter so it provides the best description for short-term variations on decade to century.

\subsection{Issues of time keeping in brief}

The time used in the dynamical equations of the ephemeris model is reckoned in a relativistic (coordinate) timescale, expressed in TDB (time 
dynamical barycentrical) units. This TDB time units evolves in a different scale with respect to the timescale in which our civil time (the time 
expressed in our watches) is based: the UTC (universal time coordinate). Our civil time, can be easily related to true solar time, which 
describes the Sun's position with respect to the observer's meridian at certain moment for a date. 
But UTC is not a dynamical timescale (i.e., UTC is not a relativistic-coordinate time). 
The corresponding coordinate time for an Earth observer is TT 
(terrestrial time). Fortunately, the difference TDB$-$TT is less than 2 ms for several millennia \citep[see e.g.,][]{Kaplan2006iau}. So we can 
 presumed TT to be equivalent to TDB (and vice versa) as a temporal reference related to orbital elements and all derived quantities from them.  
Therefore, we can assume that one can accurately represent the Earth's orbital elements in TT timescale. 
For example, a solar longitude recorded in DE431 at fiducial J2000.0 epoch, corresponds to Julian day JD 2451545.0 (TT), or the calendar date 
1/1/2000, 12 h TT, which is equivalent to 1/1/2000 11:58:55.816 UTC; i.e., at the approximate noon of an observer at Greenwich.

The relationship of UTC with TT has its own problems. UTC is based on atomic international time, which has a uniform scale, but UTC is 
subjected to leap second corrections. The corrections were done in order to keep UTC nearest to the rotational time; which is in turn non-regular
 because of the Earth's irregular rotation. The effect of 
the variability in Earth's rotation has a clear documented evidence in the past and it has a fundamental role on daily insolation determination, 
because the solar day depends on Earth's rotation. Through this leap second correction, UTC is closely related to the universal time 
(UT). This timescale has all the irregularities of Earth's rotation. UT is studied and adjusted using ancient astronomical observations and 
following \cite{Morrison2004}  for the interval 9999 BC-700 BC, the difference TT$-$UT (which, for our purposes we can assume as TDB$-$UT) 
follows a parabolic relationship such that for BC10000 = 1/1/-9999 = JD$-$1931076.0 (TT), we have TT$-$UT = 5.17 h, which is often referred to as 
the ``clock error". 
This difference shows us that a certain phenomenon (e.g., the Sun's position at noon) reckoned in TDB relativistic-uniform timescale, has 
probably an uncertainty of that amount when we try to refer as a geocentric phenomenon to be observed at that epoch. This also implies that the 
present 
nominal 86400 s for the length of the day (LOD) is no longer valid for ancient ages. This clock error is not usually mentioned nor discussed in 
paleoclimatic studies, but it is important to take into account for detailed modeling into the past. Therefore, the daily solar irradiance that 
has been extensively used \citep{Berger1988,Berger2010}, based on LOD of 86400 s, produces a very different estimation of the ``real" daily 
insolation received in a whole true solar day of a certain remote epoch in the past (or future). Regarding civil calendars, it is important 
to remember that calendar days refer to the Julian calendar dates before 4th October 1582. Julian calendar has its own reckoning of the year 
duration; then, the JD for some events in the past (as solstices or equinoxes) has an offset of several calendar months with respect to the 
present calendar.

\subsection{Obtaining Earth's orbital elements from DE431}

The JPL website ({\tt ftp://ssd.jpl.nasa.gov/pub/eph/planets/Linux/de431}) provides a binary file of DE431 ephemeris, 
 tailored for Lynux users, the {\tt lnxm13000p1700.b} file; the temporal limits of this file are 9/12/-13001 and 11/1/17000.  As far as we know 
there is no simple program (e.g., FORTRAN code) to read and directly obtain $\mathbf{r}$ and $\mathbf{v}$ from it. 
But JPL provides a Fortran code for testing another version of ephemeris in binary blocks, {\tt THESTHEP.f} routine. 
We have modified this routine to get Earth's $\mathbf{r}$ and $\mathbf{v}$ with full precision for an arbitrary time step.
Then, we performed an orbital calculation with a time step at every one Julian day. Thus, we are getting
 astro-climatic elements at each day which is a huge improvement over other estimates which considered only annual-resolution settings 
\citep[e.g.,][]{Loutre1992}. In this way we obtain a full set of Earth's orbital elements and positional angles: the semi-major axis, 
eccentricity, inclination, 
longitude of the perihelion, longitude of the ascending node, mean anomaly, true longitude and the true anomaly all with respect to the reference 
epoch J2000.0, in TDB $\simeq$ TT timescale. Therefore, for present times or instances, assuming a constant Earth rotation speed (86400 s), 
an observer at Greenwich meridian has such a precise set of Earth's elements and solar longitudes, at noon, to calculate insolation 
quantities. 

In what follows, and mainly to represent quantities as function of the time, when we make mention of  ``time in years'',  
for a specific Julian day number $J$, we refer to Julian epoch:

\beq
t = 2000.0 + \frac{J - 2451545.0}{365.25}, 
\eeq

\no which enable us to reckon the variable time, $t$, as a real number (i.e., in years and the fraction of a year).

The orbital calculation only determines $\varpi$ and $\lambda$ measured from the fixed fiducial ($\gamma_0$) J2000.0 mean equinox and ecliptic. 
However, in order to obtain these quantities but measured with respect to the true equinox of the corresponding date, we 
must deal with the long-term and short-term motions of both the equator and the ecliptic.

\section{The motion of basic planes}

Luni-solar and planetary torques modify the Earth's celestial pole and the pole of the ecliptic, which are described by the combined actions of 
the longer term (i.e., longer than 100 centuries) precessional as well as the shorter nutational changes. 
Mean precessional changes describe the evolution of the mean equator and ecliptic at a certain date. 
In our STOF calculation, we want to take into account the shortest variations of these planes;   i.e., we need to describe both the 
{\it true} ecliptic and equator of the date. 
The ecliptic is defined as the mean plane of the Earth-Moon barycenter orbit around the SS 
barycenter (i.e., perpendicular to the Earth-Moon orbital angular momentum vector with respect to the SS barycenter). Both, the ecliptic and 
equatorial planes, are subjected to luni-solar and planetary perturbations \citep{Vondrak2011,Capitaine2015}. 
Fig. 3 shows the precessional elements with respect to the J2000.0 reference epoch. The accumulated  precession (subscript A) of the equator with 
respect 
to the reference epoch is described by $\psi_A$ and $\omega_A$ (in longitude and obliquity, respectively); the corresponding elements of the 
ecliptic precession are $\Pi_A$ and $\pi_A$. The mean obliquity of the date (i.e., only affected by precession) is $\epsilon_A$. 
Then, the {\it general precession in longitude} accumulated from J2000.0 is:

\beq
p_A = \Lambda_A - \Pi_A,
\eeq

\no an arc of great circle describing the absolute motion of the mean $\gamma$ point along the ecliptic of date; it combines the 
precession in longitude of equator and the precession of ecliptic. To represent the long-term variations of these elements we used the model 
provided in \cite{Vondrak2011}, which is a significant improvement with respect to the model described in \cite{Loutre1992} that was in turn 
based on the older IAU standard of \cite{Lieske1977}. \cite{Vondrak2011}'s parametrization is based on numerical representations of mean 
equator and ecliptic \cite[e.g.,][]{Laskar1993}, which are consistent with the JPL's Development Ephemeris.

\subsection{Combining secular and short-term variations for new astro-climatic elements}

The  elements coming from DE431 define and orient the plane of the Earth's orbit of the date with respect to the reference epoch (J2000.0). 
Recall the 
fact  that DE431 separates the Earth from the Moon, therefore we can describe the ``true" Earth's orbit around the Sun. 
Then, for our astro-climatic short-term solution, we need to combine the long-term/short-term variations represented by precessional-nutational
parameters and those specific Earth's orbital orientation elements which contain both secular and short-term variations that in turn 
were derived from the DE431 ephemeris.
 This particular geometry is described and 
summarized in Fig. 4. Although DE431 uses an improved precession model for the orientation of Earth; which is a rigorous numerical integration of 
the equations of motion of the celestial pole using Kinoshita's solid Earth model \citep{Kinoshita1977} for the speed of luni-solar precession 
\citep{OwenJr1990}, which, however, is not the actual \cite{Vondrak2011}'s model. We think that such a mismatch is not a big drawback because 
the impact of Earth's true orientation on Earth's center of mass integration should be small for the Holocene time-span we considered in this 
paper. We contacted Dr William Folkner of NASA JPL and he strongly recommended using \cite{Vondrak2011} precession formulas as the detailed 
model for the long-term evolution of Earth's orientation to couple jointly to the DE431 ephemeris.  

In Fig. 4, we replace the ecliptic of the day with the Earth's orbital plane, defined through DE431, so we approximate:

\beq
\pi_A = i;  \ \ \Pi_A = \Omega,
\eeq

\no where $i$ is the {\it inclination} of the Earth's orbit in the fiducial reference system; $\Omega$ is the ascending node.
In addition, we add the nutation in longitude and obliquity to the equator movement, therefore:

\beq
\psi_t = \psi_A + \Delta \psi_A;  \ \ \omega_t = \omega_A + \Delta \omega_A,
\eeq

\no where $\psi_A$ and $\omega_a$ are obtained from \cite{Vondrak2011}'s theory. The $\Delta \psi_A$ and $\Delta \omega_A$ angles are the 
nutation in longitude and obliquity. The IAU nutation model, based on the 1980 IAU theory, is now being replaced with the new model of 
 \cite{Mathews2002}. Fortunately, its 
computational implementation is provided in the Standards of Fundamental Astronomy \citep{SOFA2016} through the SOFA software. We used IAU2000A 
model, which is the most complete version that includes all planetary and luni-solar effects on a non-rigid Earth with multiple couplings 
\citep[see][]{Mathews2002}.

At this juncture, it is important to note that the arc describing the moving equinox ($\gamma$) over the orbital plane of the Earth is $p_t$.
 We could call $p_t$ ``the true general precession of the date'' because it is the general precession affected by nutation and also by the 
approximate Earth's orbit as specified from DE431 ephemeris. Therefore, the arc $N\gamma$ is now $p_t+\Omega$ in this new description 
(see Fig. 4).
 Hence, the changes in orbital longitude and in the longitude of the perihelion, described in Section (\ref{Sec3}), should be written as:

\beq
\varpi_t = \varpi + p_t; \lambda_t = \lambda+ p_t,
\eeq

\no and this is equivalent to a change in the origin of longitudes from $\gamma_0$ to $\gamma$ in Fig. 2. 
Also the obliquity in Eq. (\ref{rdelta}), must be replaced with the ``true 
obliquity of the date'', $\epsilon_t$.  This permits us to find a new 
astro-climatic solution that we called the DE431-IAU solution. 

\section{New STOF astro-climatic elements: DE431-IAU solution}
\label{Sec6}

Now, we must find $p_t$ and $\epsilon_t$ values as a function of time to obtain the true longitudes and obliquity of the date.  Consequently, we 
need to solve 
the spherical triangle of Fig. 4, for which we follow a similar procedure as in \cite{Loutre1992}.  Using cosine formula (by angle):

\beq
\label{cos}
\cos \epsilon_t = \cos \omega_t \, \cos i - \sin \omega_t \,  \sin i \, \cos(\psi_t+\Omega)
\eeq

\no and sine formula: 

\beq
\sin(p_t+\Omega) \sin \epsilon_t = \sin \omega_t  \, \sin(\psi_t + \Omega)
\eeq

\no and five elements formula \citep[2 sides, 3 angles; see e.g.,][]{Duriez2002}:

\beq
\cos(p_t+\Omega) \, \sin \epsilon_t = \cos \omega_t \, \sin i \, + \, \sin \omega_t \cos i \, \cos(\psi_t+\Omega)
\eeq

\no we get:

\beq
\label{tan}
\tan(p_t+\Omega) = \frac{\sin \omega_t  \, \sin(\psi_t + \Omega)}{\cos \omega_t \, \sin i \, + \, \sin \omega_t \cos i \, \cos(\psi_t+\Omega)}.
\eeq

We wish to note a typographical error in \cite{Loutre1992}'s Eq. (12): the second member on the right hand side should be ``+".  
By solving Eqs. (\ref{cos}) and (\ref{tan}) we find a new set of astro-climatic parameters. 

As a matter of consistency, we compare our results ($p_t$ and $\epsilon_t$); but removing the effects of nutation correction, with the 
corresponding long-term precessional elements of \cite{Vondrak2011}, $p_A$ and $\epsilon_A$. The differences are smaller than 30 arcsec in 
obliquity and smaller than 60 arcsec in longitude for the whole 13 kyr interval studied; 
which produce a very small change in the solar radiation received, 
of $\sim 2\times10^{-4}$ W m$^{-2}$ at a maximum. 
Such a difference is significantly 
smaller than the nominal error committed in the theoretical calculation of daily irradiance (see next section). 
Therefore, we argue that both DE431 elements and IAU related precession+nutation models are sufficiently consistent and accurate for our purposes.

We show the full set of DE431-IAU astro-climatic elements (including {\it climatic precession}, $e \, \sin \varpi$) 
 over the full time interval in Figs. 5, 6, 7 and 8, where one point for every 120-day has been plotted to avoid over-sampling effects. 
 We wish to note that these sampling effects can produce the apparent “spurious” spikes in the representation of these elements, when you select 
these data at a greater time step. 
Indeed, the smallest periodicities involved in these orbital elements are of the order of a few months. The orbital periods of Mercury and Venus, 
which produce detectable spectral peaks in e.g., eccentricity, are about 0.24 yr and 0.6 yr, respectively (see also next section). At first 
sight, and comparing with \cite{Loutre1992} solution, the DE431-IAU solution seems to be ``richer", with the evidence being most obvious using 
the finest time resolution. The spectral analyses performed on them confirm our claim.

\subsection{Short-term periodicities detected}

We have explored spectral analyses 
on the DE431-IAU solution by 
using several methods; mainly, Lomb periodogram,
 maximum entropy and multitaper method (MTM), and have found a profuse amount of spectral lines which basically include all the 
periodicities found in previous works.
Short-term periodicities in astro-climatic elements were reported by \cite{Borisenkov1983}  studying daily insolation variations at 
several latitudes; they were: 2.7 yr, 4.0 yr, 5.9 yr, 11.9 and 18.6 yr. 
\cite{Loutre1992} determined several short-term periodicities by analyzing first the variations of the astro-climatic elements. For daily 
insolation at 65$^{\circ}$N, they found essentially the same periodicities (showing that the most important spectral peaks occur at decadal, 
say $<$ 40 yr, time-scale), plus 8.1 yr, 15.7 yr and 29 yr periods; arguing that these last 
were coming from perturbations by Venus, Mars and Saturn. These are the main periodicities detected, but longer term periodicities (e.g., $
\sim$ 40, 60, 800, 900 yr) were also detected, but needing more detailed scrutiny and careful examination. 
We agree with \cite{Loutre1992} that MTM is a good choice for analyzing astro-climatic elements because this method produces less noisy spectrum 
with an easy way to calculate confidence levels; in this regard, we have used SSA-MTM toolkit from \cite{Ghil2002}. 
Our results show strong non-linear trends in the spectra (which 
is clear from the element's graphics, Fig. 5-8), even while analyzing shorter portions of the time series; therefore,  
our analysis include a prior detrending of the time series. This is essential for a proper determination 
of the spectral peaks corresponding to the longer periods. 

In astro-climatic elements, these short periodicities are associated with planetary mean motions or combinations of them 
because the Earth's disturbing function depends on these orbital configurations \citep[see ][]{Roy1978,MurrayDermott1999}. These are generically 
called {\it inequalities} because the mean motion relationships $p\,n_i-q\,n_o$, where $p$ and $q$ are integers and $n_i$, $n_o$ are the {\it 
mean motion} of both planets, oscillates around a small (say, few degrees per year) value, but are not exactly zero (the planets are not in mean 
motion resonance; in fact, there are no exact mean motion resonances between planets in the SS). If the associated coefficient ($C_j$) in the 
periodic term of the disturbing function is significant, 
then that term of the perturbation is important. For example, if only two planets have been 
taken into account, the periodic variations of Earth's orbital elements depend on terms of the following form: 

\beq
\label{pert}
\sum_j \frac{C_j}{\,pn_i - q \,n_o} \, \sin ((p\,n_i - q \,n_o) \ t +B_j)
\eeq

\no where $B_j$ is another coefficient depending on angular Keplerian elements. Hence, if an inequality is small and the amplitude $C_j$ is 
important, we are in the presence of a significant perturbation, whose period is:

\beq
{\rm Per}(p\, n_i  - q \, n_o) = \frac{2\pi}{{\rm ABS}(p\, n_i  - q \, n_o)} = {\rm ABS}\left(\frac{1}{\frac{p}{P_i} - \frac{q}{P_o}}\right)
\eeq

\no where $P_i$ and $P_o$ are the orbital periods involved (note that $p=q=1$ corresponds to the synodic periods of the pair of planets); ABS is 
the absolute value function.  The smaller 
the significant inequality, the longer the associated perturbation period. 
In the SS the most important interaction is the Jupiter-Saturn (2:5) relationship, i.e., $ 2 n_j -5 n_s \simeq -0.4 \frac{{\rm deg.}}{{\rm yr}}$, 
which leads to a Great Inequality of period $\sim$ 800-1100 yr \citep[see e.g.,][]{Wilson1985}; which was detected by \cite{Loutre1992} in 
 the evolution of eccentricity.

In order to identify the planetary origin of these short periodicities, we need to evaluate the most important terms that produce detectable 
peaks in the spectral analysis of the Earth's disturbing function, from Venus to, at least, Saturn. The simplest and most complete way in 
achieving this is to perform the  high precision numerical simulations as provided by the 
{\tt MERCURY} software package for orbital dynamics \citep{Chambers1999} and then 
evaluate if any spectral peaks may be present 
\citep[for a deep theoretical description and specific calculation of perturbations for a couple of planets the reader is directed to][]
{Simon1987}. 
First, we integrate the Earth's orbital motions adding only one planet at a time from Mercury to 
Saturn, to identify the ``pure" periodicity of relevant perturbations; then we integrate the Earth with pairs of ``related" planets (Venus + 
Mars; Jupiter + Saturn; etc.) to see the modification to these ``pure" periodicities, and then all the relevant bodies (from Venus to Saturn) to 
see what, if any or all, of these short-term periodicities  ``are surviving" despite the predatory effects from stronger combined perturbations. 
By adopting this procedure we report in Fig. 9 that the prominent peaks related to planet Venus  are (in the order of importance, i.e., spectral 
power):\\

3.98 yr = Per(2V-3E); 2.67 yr = Per(V-2E); 8.10 yr = Per(3V-5E); 7.84  yr = Per(5V-8E),\\

\no with V and E denote the mean motions of Venus and Earth. For the planet Mars, the most conspicuous peak is:\\

15.76 yr = Per(E-2Ma),\\

\no where Ma denotes the mean motion of Mars. In Fig. 9, the minor peaks labeled as P, P2 and P3,
 but with a confidence level less than 95\%, come from Mars and correspond to the following periods:\\

2.47 yr = Per(2E-3Ma);  2.9 yr = Per(3E-5Ma); 5.25 yr =  Per( 3(E-2Ma)).\\

Peaks provided by Jupiter and Saturn mean motions (Per(J) = 11.86 yr; Per(S) = 29.4 yr) are also clearly visible. Of course the peaks around 
6 yr and 14.7 yr are the second harmonics of J and S. 
The result of these short-term simulations are in good agreement with main result of previous works. 
It is important to note that the magnitude of the power spectrum cannot be taken as a direct indication of the perturbation intensity. Only an 
approximate estimation can be deduced from the raw data; the spectrum depends on detrending, sampling, etc., hence the relative intensity of two 
peaks can largely vary for the same phenomena, if the input data are processed straightforwardly. 

Fig. 10 shows our MTM analysis (with a parabolic detrending of the raw, original data shown in Fig. 5) of the 13 kyr
 evolution of the record of eccentricity. 
All the peaks shown in Fig. 9 are detected in the detrended time series. 
Particularly, some peaks, such as the P, P2 and P3 are reinforced;
 raising above the 99\% confidence level (note that P2 is immersed in the power of the signal at around 2.7 yr).
Some higher harmonics such as 4J, 3U (i.e., forth and third harmonics of Jupiter and Uranus mean motions, respectively); P4 (the sixth harmonics 
of 8E-15Ma, $\sim$ 41 yr); and P5 (fourth harmonics of the synodic period of Saturn and Neptune of $\sim$ 36 yr), also appear. 
In particular, the perturbation with periodicities at around 3 yr strongly disturbs the (V-2E) periodicity and shifts this peak to 2.71 yr. 
Longitude of 
the perihelion or climatic precession shows basically the same periodicities (not shown). 
Notably, a spectral peak related to 4E-7Ma, not previously shown, is strongly present at around 3.56 yr.  
The other observed peaks are harmonics of the orbital frequencies present or non-linear combinations 
of the physical parameters' frequencies intervening in the definition of eccentricity.

Fig. 11 shows the MTM spectrum of obliquity from the detrended data record. Although a cubic-parabolic detrending have been applied to the raw 
series shown originally in Fig. 8, a secular trend in the spectral power is still present. This spectrum shows some common periodicities with
 the eccentricity and the precession of the perihelion signals. Periods directly coming from nutation are clearly detected, 
especially the period of lunar node ($M_\Omega$), and the inequality 2E-3Ma, which comes from the direct part of planetary perturbation on 
Earth equatorial bulge \citep{Souchay1996,Souchay1997,Souchay1999}. Other direct planetary effects on nutation come from 4E-7Ma, and 3E-5Ma. 
In particular, a periodicity of 8.85 yr is clearly visible that is related to the second harmonic of Saturn-Uranus synodic frequency 
(period of about 17.9 yr), which produces considerable torque in the SS \citep[see][]{CioncoAbuin2016}. 
Nevertheless, this 8.85 yr periodicity can also be related to the small
 perturbation of the Moon on the second order terrestrial potential $J_3$ (rigid Earth),  with the argument $-l_M + F + M_{\Omega}$; where $l_M$ 
is the mean anomaly of the Moon and $F = L_M-M_{\Omega}$, where $L_M$ is the mean orbital longitude of the Moon. 

 In summary, all of the above-mentioned spectral peaks are by far the most important periodicities lesser than one century. 
Our solution DE431-IAU is more complete in covering all the periodicities that were previously known and shown; and 
some of these periods are expressed in daily insolation quantities which in turn should be studied for any relations to weather and climate 
as well as all the underlying physical processes.

\section{Daily irradiance calculations}

The daily irradiation or daily insolation, $Q$, for a whole rotational day of duration $\tau$ and the mean 
daily irradiance, $W$, at a given latitude, can be obtained
 by using Eq. (\ref{dqdt}), integrating over one full Earth's rotation, being the 
integral 
that is only significant when the Sun is over the horizon.  In addition if we assume elements others than $H$ to be unchanging over a 
whole day, the mean daily irradiance is:

\begin{eqnarray}
\label{intW}
W  & = & \frac{1}{2\pi} \int_{-\pi}^{\pi} I(\phi, \delta, r, H) \ {\rm d}H \nonumber \\ 
 & = & {\rm TSI}_0 \  \left ( \frac{1 {\rm au}}{r}\right)^2  \frac{2}{2\pi} 
\int_{0}^{H_s}  (\sin \phi \sin\delta + \cos \phi \cos \delta \cos H ) \ {\rm d}H,
\end{eqnarray}

\no where $H_s$ is the sunset hour angle. Finally:

\beq
\label{W}
W=  
\frac{\rm{TSI}_0}{\pi} \  \left ( \frac{1 \rm{au}}{r}\right)^2  
  ( H_s \ \sin \phi \sin\delta + \cos \phi \cos \delta \sin H_s),
\eeq

\no reckoned in the same units as TSI$_0$ (W m$^{-2}$). In addition, Eq. (\ref{dqdt}) permits one to find the daily amount of solar irradiance  
between the time of sunrise ($t_s$) and sunset ($t_p$) which is:

\beq
\label{intQ}
Q =  
{\rm TSI}_0 \  \left ( \frac{1{\rm au}}{r}\right)^2  
\int_{ts}^{tp}  (\sin \phi \sin\delta + \cos \phi \cos \delta \cos H ) \ {\rm d}t,
\eeq

\no if we assume $t$ to be the mean solar-time timescale (i.e., we drop the difference with true solar time); using the rotational relationship 
with $H$ (1 mean day of $\tau$ seconds):

\beq
H = \frac{2\pi}{\tau} (t-t_o) ,
\eeq

\no then:

\beq
Q =  \frac{\tau}{\pi} 
{\rm TSI}_0 \  \left ( \frac{1 {\rm au}}{r}\right)^2  
\int_{0}^{H_s}  (\sin \phi \sin\delta + \cos \phi \cos \delta \cos H ) \ {\rm d}H,
\eeq

\no for one solar day of $\tau$ = 86400 s; and assuming the constancy of the Sun declination, $Q$ becomes:

\beq
\label{Q}
Q =  \frac{86400}{\pi} \ {\rm TSI}_0 \left (\frac{1 {\rm au}}{r}\right)^2 (H_s \sin \phi \sin {\delta} + \cos \phi \cos{\delta}\sin H_s), 
\eeq

\no which is expressed in J m$^{-2}$, if TSI$_0$ is in W m$^{-2}$. Eq. (\ref{cosz}) permits one to find the value of the $H$ at sunrise-sunset. 
When the Sun is on the horizon (i.e., Sun at 90 deg. of the zenith) the dot product is zero, hence: 

\beq
\label{D}
\cos(H_s) = - \tan \phi \ \tan \delta.
\eeq

\no Eq. (\ref{D}) acts as a discriminant for sunrise-sunset. If we call $ D = - \tan \phi \ \tan \delta$, then the Sun rises and sets at 
certain latitude when:

\beq
-1 \leq D \leq 1.
\eeq

\no If $D = -1$ for a particular observer; the Sun is at the limit of rising and setting; i.e., when $H_s = \pm 180$ deg.; the Sun becomes 
circumpolar for that observer: no sunrise-sunset for the 
time-span during $D < -1$. Of course, this occurs when $\phi$ and $\delta$ have the same sign (the Sun at the same hemispheric position). Then if:

\beq
D <-1,
\eeq

\no the Sun does not set (polar day). When the Sun's declination changes sign, the winter season arrives, and $H_s$ decreases, and for certain 
latitude, it is virtually zero; then $D = 1$. Therefore, the Sun does not rise (polar night), 
 if: 

\beq
 D> 1,
\eeq

\no and insolation quantities are null. If $D < -1$, 
then the integrand in Eq. (\ref{intW}) is significant over $0$ to $2\pi$; hence, 
the mean irradiance for the circumpolar Sun is:

\begin{eqnarray}
\label{Wc}
W_c & = &  
{\rm TSI}_0 \  \left ( \frac{1 {\rm au}}{r}\right)^2  \frac{1}{2\pi} 
\int_{0}^{2\pi}  (\sin \phi \sin\delta + \cos \phi \cos \delta \cos H ) \ {\rm d}H  \nonumber \\
 & = &
{\rm TSI}_0 \  \left ( \frac{1 {\rm au}}{r}\right)^2  \ \sin \phi \sin\delta.
\end{eqnarray}

\no Then, Eq. (\ref{intQ}) becomes:

\begin{eqnarray}
\label{Qc}
Q_c & = & {\rm TSI}_0 \  \left ( \frac{1 {\rm au}}{r}\right)^2 \frac{86400}{2\pi} \int_{0}^{2\pi}  (\sin \phi \sin\delta + \cos \phi \cos \delta 
\cos H ) \ {\rm d}H  \nonumber \\ 
 & = & \ 86400 \, {\rm TSI}_0 \  \left ( \frac{1 {\rm au}}{r}\right)^2   \  \sin \phi \sin\delta. 
\end{eqnarray}

Therefore, using Eq. (\ref{D}) as discriminant, we can calculate by means of Eqs. (\ref{W}, \ref{Q}, \ref{Wc}, \ref{Qc}), these daily solar 
quantities, knowing all of the other corresponding solar-terrestrial parameters for a certain date.  As a matter of consistency, we see that the 
averaged quantities can be directly obtained in dividing the daily values by 86400 ($W = Q/86400$). 

In the derivation of Eqs. (\ref{W}, \ref{Q}, \ref{Wc}, \ref{Qc}) we adopted a day-based approach, i.e., taking the longitude data 
tabulated (at noon) for that date. 
 This particular approach is the conventional procedure commonly used in obtaining daily solar quantities \citep[see e.g., ][]
{Berger1988,Laskar1993,Berger2010}.
Nevertheless, as we have hinted at the beginning of this article, it is an usual practise in paleoclimate 
studies, to calculate $Q$ (or $W$) by fixing the Earth/Sun position at an arbitrary moment; i.e., fixing an 
arbitrary value of solar longitude inside a particular day; this is equivalent to taking Eqs. (\ref{W}, \ref{Q}, \ref
{Wc}, \ref{Qc}) as a continuous function of longitudes.  
For example, to follow the daily irradiation at March equinox for a mid-latitude observer, the Eq. (\ref{W}) is used by 
setting $\lambda_{\odot \, t} = 0$ 
 as a temporal reference in Eq. (\ref{rdelta}), to find $\delta$.
This approach is understandable because at ancient 
 time we do not know these parameters ``exactly'' at noon.
Moreover, we can assume that the 
change of solar longitude inside a day does not significantly affect the estimate of daily insolation \citep{Berger2010}. 
Nevertheless, we consider the clarification and recognition of this assumption to be important for all users of available databases of solar 
irradiation (see next Sub-Section 7\ref{Errors}).

Fig. 12 shows the daily mean irradiation at equinoxes ($\lambda_{\odot \, t} = 0, 180$ deg.), using our DE431-IAU solution, over the indicated 
 calendar time, at 65$^{\circ}$N. 
 The solar constant used is 1366 W m$^{-2}$. Fig. 13 shows similar quantities as in Fig. 12 but for solstices; i.e., $\lambda_{\odot 
\, t} = 90, 270$ deg. 
Both calculations were performed taking Eq. (\ref{W}) as a continuous function of solar longitude. As were pointed out by \cite{Borisenkov1983} 
and \cite{Loutre1992}, the secular long-term modulation by precession is evident for equinoxes, even at high latitudes. 
For solstices the dominant signal is provided by the modulation of the obliquity variations. 
 For daily insolation calculations, we prepare a {\tt FORTRAN} code that calculates $Q$ and $W$ (Eqs. \ref{W}, \ref{Q}, \ref{Wc}, \ref
{Qc}) along the time for a specific solar longitude and latitude on Earth. 
 The code {\tt INSOLA-Q-W.for} permits the calculation of $Q$ or 
$W$ for the exact mid-day at the tabulated longitude of a certain Julian day, or to set $Q$ or $W$ as continuous functions of the solar 
longitude, which is the common practice in paleoclimate studies. Fig. 12 and 13 were made with this last strategy, i.e., by setting  $\lambda_
{\odot \, t}$ with the desired value (0, 90, 180 or 270 deg.) using Eq. (\ref{W}) and (\ref{Q}), but taking the DE431-IAU elements for the 
specific tabulated JD within which the corresponding value, in our cases of equinoxes or solstices, occurs. Hence, these values were calculated 
for the specific astro-climatic parameters of the corresponding mid-day, but did not used the tabulated mid-day longitude, as daily insolation 
should be calculated in theory. The difference between the desired longitude and the tabulated mid-day value can reach $\sim \ \pm$ 0.7 deg.

Fig. 14 shows $W$ at the ``mid-month" of July; i.e., $\lambda_{\odot \, t}$ = 120 deg.
 In keeping with the usual practice in paleoclimate, we are continuing to adopt the present calendar defining  mid-month moments at an angular 
step of 30 deg., from March (0 deg.) to February (330 deg.). 
Then the calculation was also performed taking $W$ as a continuous function of $\lambda_{\odot \, t}$. The period corresponds to the same 
interval considered in Figure 12a of \cite{Loutre1992}, from 1450 AD to 1950 AD. In fact, this permits a direct comparison of our result with 
\cite{Loutre1992}; and this comparison yields a very good agreement. We do not know the precise TSI$_0$ value adopted by those authors (but we 
presumed it was about 1366-1367 W m$^{-2}$ during solar activity minima as hinted on p. 191 of Loutre et al. 1992), but this level of agreement 
is reached with 1366 W m$^{-2}$. 
The obliquity signal is prominently present, with the two main periodicities clearly visible: the $\sim$ 20 yr oscillation due to 
lunar retro-gradating motion (18.6 yr) and the strong  $\sim$ 2-3 yr period.    
 
With regards to other periodicities, we performed an extensive spectral analysis
 on daily insolation outputs (using detrended time series), especially focused at 
 the shortest periodicities on biennial to 40 yr timescales (the main periods present in astro-climatic elements). 
 For a comparison with \cite{Loutre1992}, we show the results for July mid-month at 65$^{\circ}$N  (Fig. 15). 
 Several periods visible in the eccentricity, longitude of the perihelion and obliquity signals are present, 
but also prominent harmonics and non-linear interactions of frequencies present in the parameters that define daily insolation. 
The shortest periods oscillate at about 2.42 yr and 2.69 yr and correspond to the perturbed periods 
by 2E-3Ma and V-2E, respectively. 
At around 3 yr there are several harmonics and 3E-5Ma (2.9 yr), 4E-7Ma (3.56 yr), 4J;  etc. Then 3.98 yr (2V-3E); 5.26 yr (3(E-2Ma)); 
2J; 7.88 yr (5V-8E); 8.09 yr (3V-5E); Per(J);  Per(2 $M_\Omega$), Per(2S), Per($M_\Omega$)  and Per(S) complete the most important periods 
present.  
For lower latitudes (e.g., 0 deg.), the spectrum is essentially the same; the only change is the attenuation of the obliquity signal; as expected 
because precession signal is dominant at tropical latitudes. At these lower latitudes, the 18.6 yr period and its harmonics appears smaller than, 
for example, the Jupiter signal. 

Although our derived periodicities are similar to \cite{Loutre1992}, we found a richer set of periodicities at around 2-3 yr, 5 yr, and 8 
yr, coming from a more complete description of planetary perturbations performed in DE431 and also the more complete model of nutation.
Also, by performing a much more detailed detrending and sampling, we can also produce similar results as those
 discussed in \cite{Loutre1992} for the longest periodicities covering oscillations on timescales from multicenturies to a millennium. 

In addition, we have created another code {\tt MONTH-GEO.for} designing to calculate the geographical distribution of mid-month insolation for 
arbitrary 
years (between $-$10 kyr, 3 kyr), in different latitudinal bands. An output for the year 1650, near the beginning of the Maunder Minimum, can be 
seen in Table 1, for the indicated latitudes. This code calculates $W$, as continuous functions of $\lambda_{\odot \,t}$, i.e., 
at the corresponding longitudes for each mid-month of the conventional calendar. 
All our codes are available with the publication of this paper.

\subsection{The effect of the non-constant longitude within a day}
\label{Errors}

As discussed in details in the previous section, $Q$ and $W$ formulas were derived assuming that we have astro-climatic elements and true orbital 
longitudes
 at solar noon ($H=0$) and these values do not change significantly over a complete $2\pi$ variation of $H$. This approximation is supposed to 
lead 
to a maximum theoretical error of 0.01 W m$^{-2}$ in the determinations of $W$ \citep{Berger2010}.

In our STOF calculations, we know the value of the true solar longitude at mid-day and the astro-climatic elements for that day. Therefore, we 
can estimate the difference or ``error" in the estimation of corresponding solar quantity between the exact tabulated value 
at mid-day ($Q_0$ and $W_0$) and the estimation considering $Q$ or $W$ as a continuous function of $\lambda_{\odot \, t}$: 
the difference, $Q-Q_0$, $W-W_0$ and also the ``relative error" ratio, $(Q-Q_0)/Q_0 = (W-W_0)/W_0$. As far as we know, these differences have not 
been discussed nor addressed in the literature.

To describe these differences as a function of time, we have shown in Fig. 16 and 17 (for 65$^{\circ}$N) the relative error ratios evaluated at 
equinoxes and solstices for all the 13 kyr studied period.  
At the solstices, when the Sun is stationary, the relative errors are small. 
 Although there are values as large as 0.2\%,  they are all mostly less than 0.1\%. 
For the equinoxes 
the relative differences are 
larger, reaching $\pm$ 1.2\%. Evidently, we are confirming that the characteristic of this 
relative error is dependent on solar longitude. To explore this, we calculate this error as a function of longitude for every 10-degree 
interval. The result is shown in Fig. 18. 
We observe very large relative differences of up to $\pm$ 5\% around December solstice. 

The biggest relative differences between nominal $Q$ or $W$ values and the corresponding values as continuous functions of time, occur at the 
moment when the Sun is near the extreme declination at opposite hemispheres.
 This means that these big relative errors occur in the 
seasons when the insolation values are near minimum (of course, the same occurs for a southern observer); 
then, there are not so important 
differences on the amount of modeled received energy at these moments, as we can see in Fig. 19, where $W$ values (in continuous and tabulated 
modes) for $\lambda_{\odot \,t} = 250$ deg. (i.e., November-December), are depicted for the last 1 kyr. 
Nevertheless, as we can see in Fig. 20, this error
for the mean daily insolation,
 can increase up to $\sim$ 2.5 W m$^{-2}$ in absolute values. These largest errors occur after the March equinox 
(at the beginning of the boreal spring) and at the end of the boreal summer. 
The behavior of these differences depends strictly on the trigonometric part of Eq. (\ref{W}), which is a monotonically increasing 
function of declination, which varies approximately between $7\times10^{-3}$ and 1 between December and June, respectively, for a northern 
observer. 
The smallest errors are occurring around solstice of December because of the very small differences resulting from the subtraction between two 
smallest values of this function. Conversely, slightly larger errors with respect to December solstice, occur near solstice of June,  
when the expression for the trigonometric function in Eq. (\ref{W}) 
has its largest values.

The characteristics of this error is similar as a function time, over the full $13$ kyr interval studied. Of course, this error could lead to 
large differences  for initial conditions in a climatic modeling code especially when the climatic system is evolving nonlinearly while 
encountering 
the chaotic regime of the climate dynamic phase space.
Therefore, the difference in evaluating $Q$ or $W$ as continuous functions of longitudes or tabulated values, provides a clue of the 
``uncertainty" associated with the correct incoming solar radiation when the daily insolation value is basically a {\it guess} in connection to 
climate modeling.

\subsection{Comparison with the state of the art works by J. Laskar and colleagues}

The longest database available for Milankovit\'c orbital forcing is the one provided by \cite{Laskar2004},  and its extension to $-$250 Myr in 
\cite{Laskar2011}.  While this solution in astro-climatic elements is given for $-$250 Myr from present, inherent problems to its chaotic 
evolution make it strictly valid, in paleoclimate studies, over about $-$50 Myr and $+$20 Myr from the present, which is the time-span 
covered by former \cite{Laskar2004} solution.  
 This French solution (available at NOAA/NCDC which redirect to L'Institut de M\'ecanique C\'eleste et de Calcul des \'Eph\'em\'erides -IMCCE-) 
is consistent with \cite{BergerLoutre1991} for the last 5 Myr where latter  was, in turn, based 
on the original \cite{Laskar1986,Laskar1988} results. 
This is the reason we have considered only the solution calculated by Laskar and colleagues which we have labeled here as La2010 solution, as the 
authors named their last solution in the paper of the  year 2011 for a comparison to our DE341-IAU solution. 
The comparison also allows indirect assessments of Berger and Loutre solutions. 
The web  interface at IMCEE, permits outputs from $-$100 Myr to $+$20 Myr from J2000.0, but at a minimum timestep of 100 yr; hence this is the 
maximal resolution of this data that we took into account, which is fine enough for our purposes. 

We compare mean daily insolation at July mid-month (i.e., $\lambda_{\odot \, t} = 120$ deg.) for 65$^{\circ}$N, i.e., as a 
continuous function of longitude, using the {\tt INSOLA-Q-W.for} code. Fig. 21 shows this calculation for 12 kyr, from year $-$10000.0 to 2000.0. 
Whereas our results describe the short-term variations on $W$, La2010 solution describes the mean variations
(to make our results clearly visible at this scale, we have decided do not graph La2010 solution from $-$12 kyr to $-$10 kyr, which 
describes the mean values of our DE431-IAU solution). 
A zoom between $-$10 kyr and $-$9.6 kyr 
from J2000.0, shows the excellent agreement of our result with La2010, and the effects of STOF on $W$. To show the 
differences
 between continuous and tabulated values in our DE431-IAU solution, we performed the same calculation, using tabulated values at mid-day. 
Fig. 22 shows these differences in detail for the past 1 kyr. 
There are differences up to $\pm$ 0.3$\%$ which represents a maximum error of about 1.2 W m$^{-2}$ for this particular July mid-month 
calculation. 
Hence, a user that calculated the mean daily insolation
 using La2010 solution, can have an error up to 1.2 W m$^{-2}$ 
considering solar position at the noon time of the corresponding day, for July mid-month.

As a final commentary, \cite{Loutre1992} had shown that the level of disagreement of their solution with the original solution from Berger 
(1978) can be as large as 10 W m$^{-2}$ for 10 kyr to 5 kyr before 1950 AD; while the disagreement is more tolerable at level of less than 1 W m$^
{-2}$ for 5 kyr to present. Hence, our work presents a better agreement with respect to the state of the art solution by \cite{Laskar2011}. 
 Our STOF solutions do not show evidence for any long-term growth in comparison to the La2010 solution; La2010 describes virtually the mean value 
of our STOF solution. 
 The departure of our 
(continuously-varying-longitude) solution with respects to secular variations are indeed small, less than about $\pm$ 0.15 W m$^{-2}$ over the 
full Holocene interval covered in our DE431-IAU solution. 
 As an example, for this July comparison, the long-term La2010 value corresponding to $-1$ kyr from J2000.0 (obtained from web interface) is $W$ 
= 430.162 W m$^{-2}$, keeping three decimal figures; our DE431-IAU solution (in continuous mode), for mid-July, i.e., with $\lambda_{\odot \, t} 
= 120$ deg. 
(which correspond to $\ t$ = 1000.577 yr of our solution, with a true mid-day longitude of $\lambda_{\odot \, t} = 119.673$ deg.) give us W = 
430.238 W m$^ {-2}$, a difference of less than $-0.018$\%. 
For $-$10 kyr from J2000.0, La2010 gives W = 468.978 W m$^{-2}$, whereas our solution gives  $W$ = 468.968 W m$^{-2}$ (at $t$ = $-8000.2382$ yr), 
a difference of $\sim$ 0.002\%. 
This independent confirmation suggests that our solution can be applied for exploring any climate modeling experiments accounting fully for the 
STOF aspects of the orbital forcings previously not available publicly.

\section{Summary}

We have presented a new set of calculations accurately accounting for all short-term orbital modulations that can be considered an improved 
representation of the boundary conditions relevant for any meteorological and climatic studies over the Holocene interval and 1 kyr into the 
future. We have reviewed the subject carefully and discussed all the steps and assumptions involved in our calculations. 
 Our new orbital solutions, DE341-IAU, offer an internally self-consistent set of boundary conditions readily applicable for climate model 
simulation, attribution and even assimilation studies. The rich spatio-temporal dynamics of the persistent solar irradiation forcing offers a 
realistic accounting of all STOF dynamics suitable for the evaluation of all atmospheric, oceanic and coupled air-sea oscillations on timescales 
ranging from diurnal, fortnightly, monthly, sub-annual, seasonal, annual, QBO, ENSO, decadal to multidecadal timescales.

We offer, for the first time, a detailed determination of the error committed using daily insolation quantities
as a continuously varying longitude over the day (i.e., the most common practice in all paleoclimate modeling study thus far)  
 with respect to the nominal (based on solar position at noon) values. 
We found errors up to 5\% in daily insolation quantities, which correspond to an absolute difference of 2.5 W m$^{-2}$ 
in $W$ calculations,  
which are significantly larger 
than the theoretically expected error \citep[i.e., 0.01 W m$^{-2}$ reported in][]{Berger2010} in the calculation of $W$. 
This error is of similar order as the uncertainty detected by \cite{Joussaume1997} due to the ``calendar effects'' (i.e., 5-8 W m$^{-2}$), which 
have been shown to be considerably consequential on climate modeling \citep[see also][]{Chen2011}. 
 However, when compared to the relatively well-documented calendar effect,
it is important to note that our continuous-versus-tabulated longitude error is a different effect 
which can occur even in the same specific calendar day.    
Therefore, climate modelers have to be aware of 
these differences and how this particular error would spread and propagate along with time 
especially when intra-annual insolation quantities are required. 
The sensitivity of initial climate conditions to these differences should also be addressed.
 
Our next step will be to 
incorporate a range of estimates of intrinsic changes in TSI for the past few thousand years as constrained both by solar activity proxies from 
paleoclimatic evidence as well as observational study of solar-type stars. 
The astro-climatic parameters database, the {\tt FORTRAN} codes and all the solar radiation values will be made available to any 
scientists and climate modeling groups requesting for them. We will also make effort to install our solar radiation forcing outputs on several 
popular data archive centers including the one at NOAA/NCDC. 
Finally, after the completion of our calculation, we discover that another parallel and independent effort on tackling STOF has been ongoing in 
Russia by \cite{Fedorov2015} where the author adopted the JPL DE 405/406 Planetary and Lunar Ephemerides covering the time interval from 3000 BC 
to 2999 AD. 

%
\section*{Acknowledgments}
The authors acknowledge the support of the grant UTN-3577 ``Efectos Orbitales sobre la Irradiancia Solar Recibida Durante el Holoceno y los 
Pr\'oximos 3000 A\~nos" (2015-2016) of the Universidad Tecnol\'ogica Nacional, Argentina. 
 The authors are indebted to William Folkner of NASA-JPL, for an open and deep discussion of DE-JPL 
ephemerides models and details. 
We also extend thanks to Jean-Louis Simon of Paris Observatory-IMCEE, France, for sharing one of his papers and valuable commentaries. 
Special thanks to John Chambers of Carnegie Institution of Washington, for sharing his Julian Day routine that is especially useful for negative years.
Finally, we are grateful to Maria McEachern, the 
librarian at the John Wolbach library of the Center for Astrophysics, for valuable help.
W.S.'s work is indirectly supported by SAO grant proposal ID 000000000003010-V101. 
W.S. would also like to acknowledge the influence by Dr. Duncan Steel's independent examination of the STOF question
 and to thank the late Bob Carter, Milan Dimitrijevic, Monika Jurkovics, Laszlo Kiss, Katalin Olah, and Dmitry Sokoloff for helps and encouragements.

\bibliographystyle{elsarticle-harv} 
\bibliography{references}


\newpage
\noindent {\bf Table LIST - CAPTIONS}

Table 1: Mid-month $W$ values at indicated latitudinal bands, for the year 1650, at the beginning of the Maunder Minimum interval of reduced 
sunspot activity; TSI$_0$ = 1366.0 W m$^{-2}$ (code ${\tt MONTH-GEO.for} $). 
Each file corresponds to one month, from March to February.

\begin{table}[h!]
\begin{center}
\caption{Mid-month $W$ values at indicated latitudinal bands, for the year 1650, at the beginning of the Maunder Minimum interval of reduced 
sunspot activity; TSI$_0$ = 1366.0 W m$^{-2}$ (code ${\tt MONTH-GEO.for} $). 
Each file corresponds to one month, from March to February.}

\begin{tabular} {cccccccc} \hline \hline  
 80$^{\circ}$S &  65$^{\circ}$S  &  35$^{\circ}$S  &  0$^{\circ}$ &   35$^{\circ}$N   &   65$^{\circ}$N  &   80$^{\circ}$N    \\ \hline 
\hline        

479.093 & 457.392 & 495.965 & 421.104 & 216.929 &  16.487 &   0.000 \\
273.664 & 327.543 & 439.656 & 435.016 & 280.267 &  75.693 &   0.000 \\
 75.854 & 184.610 & 357.826 & 436.825 & 357.826 & 184.610 &  75.854 \\
  0.000 &  73.207 & 271.054 & 420.713 & 425.199 & 316.770 & 264.660 \\
  0.000 &  15.558 & 204.709 & 397.381 & 468.027 & 431.626 & 452.106 \\
  0.000 &   2.744 & 179.712 & 385.755 & 481.757 & 480.005 & 518.599 \\
  0.000 &  15.497 & 203.867 & 395.742 & 466.091 & 429.836 & 450.227 \\
  0.000 &  72.683 & 269.130 & 417.732 & 422.190 & 314.535 & 262.799 \\
 75.240 & 183.117 & 354.932 & 433.292 & 354.932 & 183.117 &  75.240 \\
271.797 & 325.287 & 436.611 & 431.991 & 278.309 &  75.155 &   0.000 \\
477.195 & 455.571 & 493.975 & 419.403 & 216.045 &  16.413 &   0.000 \\
554.521 & 513.252 & 515.122 & 412.469 & 192.155 &   2.933 &   0.000 \\

\hline
\end{tabular}
\end{center}
\end{table}


\newpage
\noindent {\bf FIGURES LIST - CAPTIONS} (ALL FIGURES IN BLACK-WHITE)\\

Fig. 1: A local fixed geocentric equatorial system $(X,Y, Z)$ is defined; $X$ directed towards the local meridian; 
 minus $Z$ axis directed towards the hemispheric pole (south, here). The minus $Y$ axis points west. 
The positions of the observer's zenith and the Sun are also indicated.\\

Fig. 2: Sketch of the Earth's orbit in the present time (reference epoch J2000.0). 
The Sun (with the usual astronomical symbol $\odot$) is at the focus. 
Nevertheless, to describe the apparent (also anti-clockwise), relative movements of the Sun with respect to Earth, we have marked the Sun 
projections on the celestial sphere, specially for the corresponding solstices and equinoxes, which permit a determination of $\lambda_{\odot}$ to be used in 
 the insolation formulas. 
Also, the effect of the moving position of equinoxes (from $\gamma_0$ of the epoch to $\gamma
$ of the date) is marked with an arrow. The dotted segment $q-A$ is the apsidal line which determines the perihelia ($q$) and aphelia ($A$). The 
direction to the ecliptic's pole is $\breve{C}$. The Earth's equator is drawn over the gray terrestrial sphere, and its continuation 
determines the Sun's declination ($\delta$) with the Sun-Earth radiovector ($\mathbf{r}$). In addition, the Earth terrestrial pole direction is 
marked, which help defines the obliquity $\epsilon$. The Earth's true longitude and the longitude of the perihelion, both of the epoch of 
reference, are also 
indicated.\\

Fig. 3: Precessional elements measured with respect to the fiducial J2000.0 epoch. The mean equator of reference (J2000.0) and the mean equator 
of the date, are indicated. The nodal point $N$ is the intersection of the two ecliptics: $E_{{\rm o}}$ of the reference epoch, and $E_t$ the 
ecliptic of the date. The general precession in longitude accumulated from J2000.0 is $p_A$ (i.e., the segment from the projection of $\gamma_0$ 
on $E_t$, to $\gamma$ point). See the main text for descriptions of other angles. \\

Fig. 4: The spherical triangle used to find the ``true general precession'' and obliquity of the date is shown. 
The $\Omega$ and $i$ elements coming from DE431 define the plane of the Earth's orbit of the date with respect to the reference epoch (J2000.0). 
See the text for more details.\\

Fig. 5: Short-term evolution of the Earth's eccentricity over 13 kyr (one datum per 120 d). The last 500 yr, from 1950, is illustrated in the 
insert with expanded details.\\

Fig. 6: Short-term evolution of the Earth's longitude of the perihelion over 13 kyr (one datum per 120 d). The last 500 yr, from 1950, is 
illustrated in the insert with expanded details.\\

Fig. 7: Short-term evolution of the Earth's climatic precession  over 13 kyr (one datum per 120 d). The last 500 yr, from 1950, is illustrated in 
the insert with expanded details.\\

Fig. 8: Short-term evolution of the Earth's obliquity over 13 kyr (one datum per 120 d). The last 500 yr, from 1950, is illustrated in the insert 
with expanded details.\\
    
Fig. 9:  MTM power spectrum of the raw data from the simulated perturbations on Earth's eccentricity by Venus to Saturn over 1 kyr (from 1000 to 
2000 AD). The labels mark the main recognized perturbations. P, P2 and P3 refer to perturbations peaks with confidence level less than 95\% 
which are related to planet Mars. 
One un-labeled peak between P2 and 2V-3E seems to be associated with the 4E-7Ma inequality. 
The other un-labeled but significant peaks are harmonics of labeled perturbations.\\

Fig. 10: MTM power spectrum for the short-term eccentricity variations based on the detrended data series after removing a parabolic long-term 
trend.  The basic periodicities identified in Fig. 9 can also be detected here.\\

Fig. 11: MTM power spectrum of the detrended obliquity time series. The main periodicities are identified (see discussion in the main text). 
Other periodicities are from harmonics and non-linear mixing of frequencies.\\

Fig. 12: Daily mean irradiation at $65^{\circ}$N over the indicated calendar interval for equinoxes; i.e., $\lambda_{\odot \, t} = 0,180$ deg. 
The solar constant used was 1366 W m$^{-2}$.\\

Fig. 13: The same as Fig. 12 but for solstices; i.e., $\lambda_{\odot \, t} = 90,270$ deg. The solar constant used was 1366 W m$^{-2}$.\\

Fig. 14: $W$ at July mid-month; i.e., $\lambda_{\odot \, t}$ = 120 deg., for $65^{\circ}$N. The time interval corresponds closest to the one  
considered in \cite{Loutre1992}'s Fig. 12a. This comparison shows a very good agreement with their results. Solar constant used was 1366 W m$^
{-2}$.\\

Fig. 15: MTM spectrum of $W$ (over detrended time series) at July mid-month, $65^{\circ}$N. The main periodicities are identified and are 
described in the main text. At lower 
latitudes, the periodicities  arising from the obliquity signals are weakened.\\

Fig. 16: Relative difference in $W$ calculation as a continuous $\lambda_{\odot \, t}$ or using tabulated values at mid-day, for solstices 
($\lambda_{\odot \, t}$ = 90, 270 deg.). 
At these moments, when the Sun is stationary, the errors are small, but 
they can reach 0.2\%. \\

Fig. 17: The same as Fig. 16 but for  equinoxes ($\lambda_{\odot \, t}$ = 0, 180 deg.). The error is larger
on average
 by one order of magnitude (or about 1.2 \%) when compared to the errors during solstices.\\

Fig. 18: Relative difference in the $W$ calculation as a function of $\lambda_{\odot \, t}$ over the 13 kyr interval we studied (for 65$^{\circ}
$N). 
Around the solstice of December, the relative error can reach 5\% but drop to the level of less than 0.1\% at 270 deg. 
\\

Fig. 19: $W$ calculation as a continuous $\lambda_{\odot \, t}$ (lines) or using tabulated values at mid-day (dots), for $\lambda_{\odot \, t}$= 
250 deg. (for 65$^{\circ}$N), from 1000 to 2000 AD. The absolute difference is less than 0.3 W m$^{-2}$.\\

Fig. 20: Difference $W-W_0$ in mean daily insolation calculation 
as a function of $\lambda_{\odot \, t}$ for the 13 kyr interval we studied (for 65$^{\circ}$N). At 
boreal spring and at the end of the boreal summer, the absolute difference can 
reach 2.5 W m$^{-2}$ (see the text for more explanation). \\

Fig. 21: $W$ at July mid-month ($\lambda_{\odot \, t}$ = 120 deg.), for 65$^{\circ}$N: this work, DE431-IAU (dashed lines); Laskar et al. (2011a) 
solution, La2010 (points). The comparison is shown  starting at $-$10 kyr with expanded details of the intercomparison illustrated in the insert.
The solar constant used was 1366.0 W m$^{-2}$.\\

Fig. 22: $W$ at July mid-month ($\lambda_{\odot \, t}$ = 120 deg.), for 65 $^{\circ}$N: DE431-IAU solution, continuous values (dashed lines); 
La2010 data (points) and DE431-IAU solution, with tabulated values (dots). Differences between the continuous and 
tabulated values can reach $\pm$ 0.3\%; i.e., approximately $\pm$ 1.2 W m$^{-2}$.\\

\newpage
\clearpage
\begin{figure}
\begin{center}
\includegraphics[scale=1]{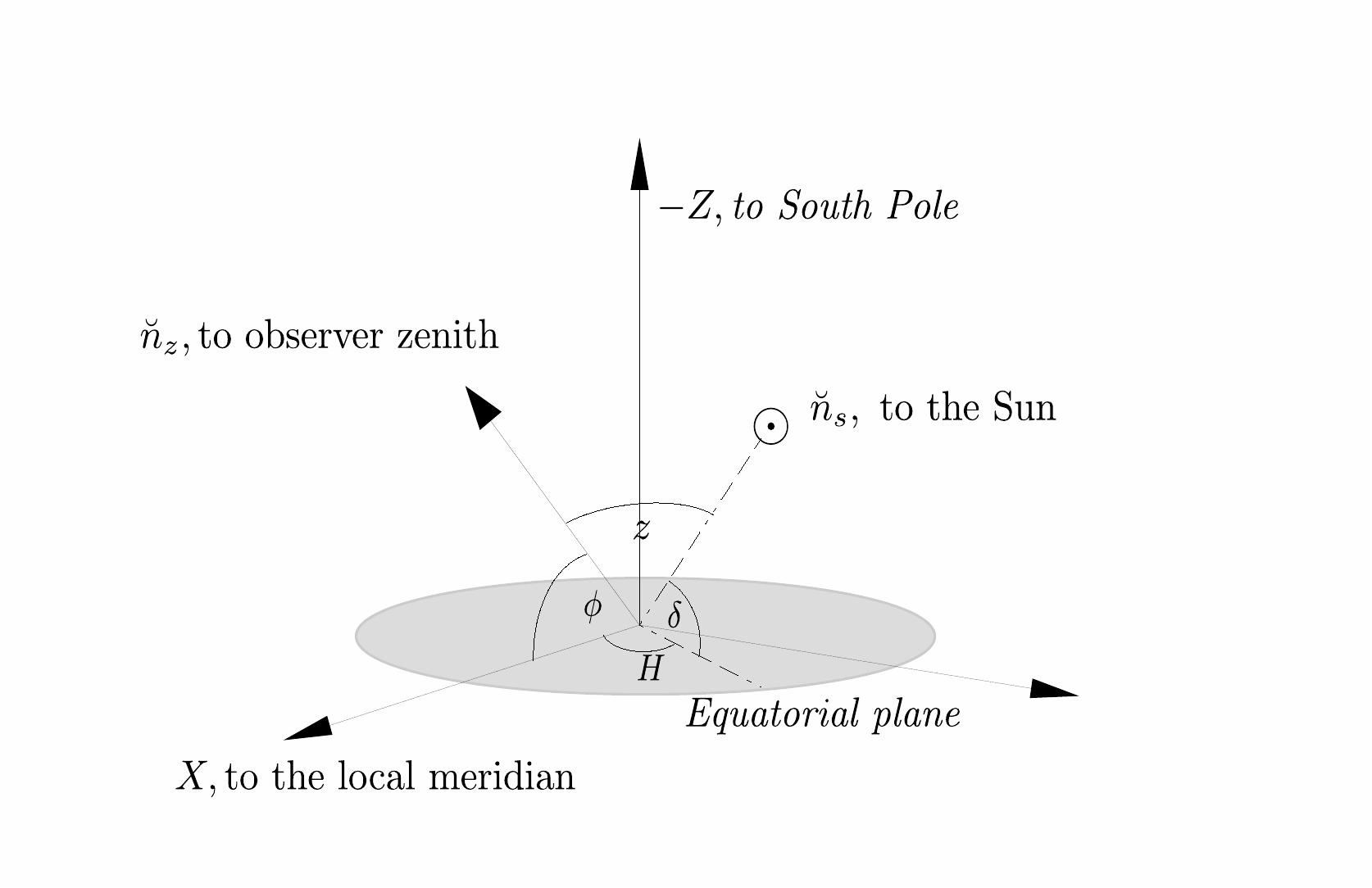}
\caption{A local fixed geocentric equatorial system $(X,Y, Z)$ is defined; $X$ directed towards the local meridian; 
 minus $Z$ axis directed towards the hemispheric pole (south, here). The minus $Y$ axis points west. 
The positions of the observer's zenith and the Sun are also indicated.}
\end{center}
\end{figure}

\begin{figure}
\begin{center}
\includegraphics[scale=0.8]{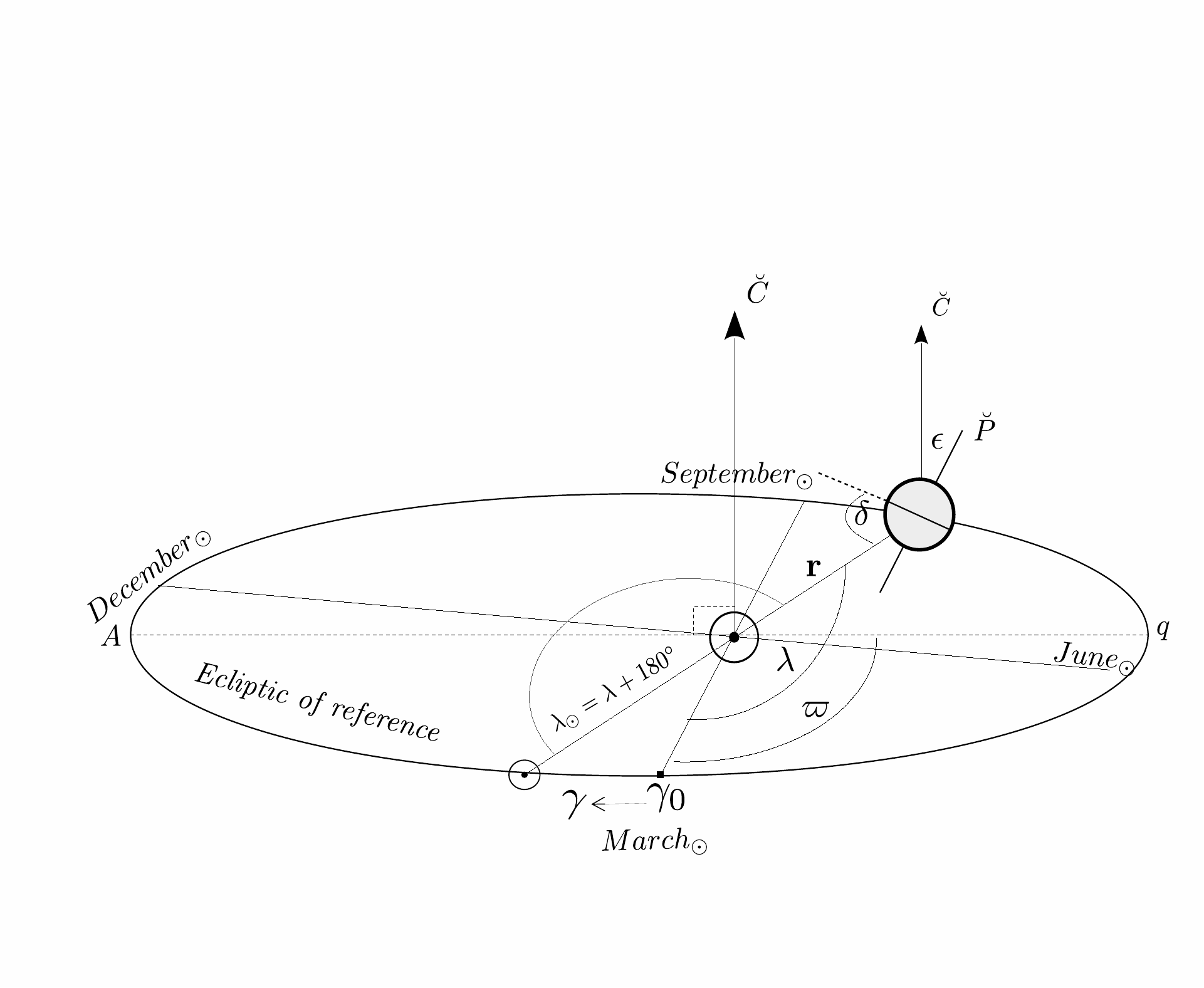}
\caption{Sketch of the Earth's orbit in the present time (reference epoch J2000.0). 
The Sun (with the usual astronomical symbol $\odot$) is at the focus. 
Nevertheless, to describe the apparent (also anti-clockwise), relative movements of the Sun with respect to Earth, we have marked the Sun 
projections on the celestial sphere, specially for the corresponding solstices and equinoxes, which permit a determination of $\lambda_{\odot}$ to be used 
in the insolation formulas. 
Also, the effect of the moving position of equinoxes (from $\gamma_0$ of the epoch to $\gamma
$ of the date) is marked with an arrow. The dotted segment $q-A$ is the apsidal line which determines the perihelia ($q$) and aphelia ($A$). The 
direction to the ecliptic's pole is $\breve{C}$. The Earth's equator is drawn over the gray terrestrial sphere, and its continuation 
determines the Sun's declination ($\delta$) with the Sun-Earth radiovector ($\mathbf{r}$). In addition, the Earth terrestrial pole direction is 
marked, which help defines the obliquity $\epsilon$. The Earth's true longitude and the longitude of the perihelion, both of the epoch of 
reference, are also 
indicated.}
\end{center}
\end{figure}

\begin{figure}
\begin{center}
\includegraphics[scale=1]{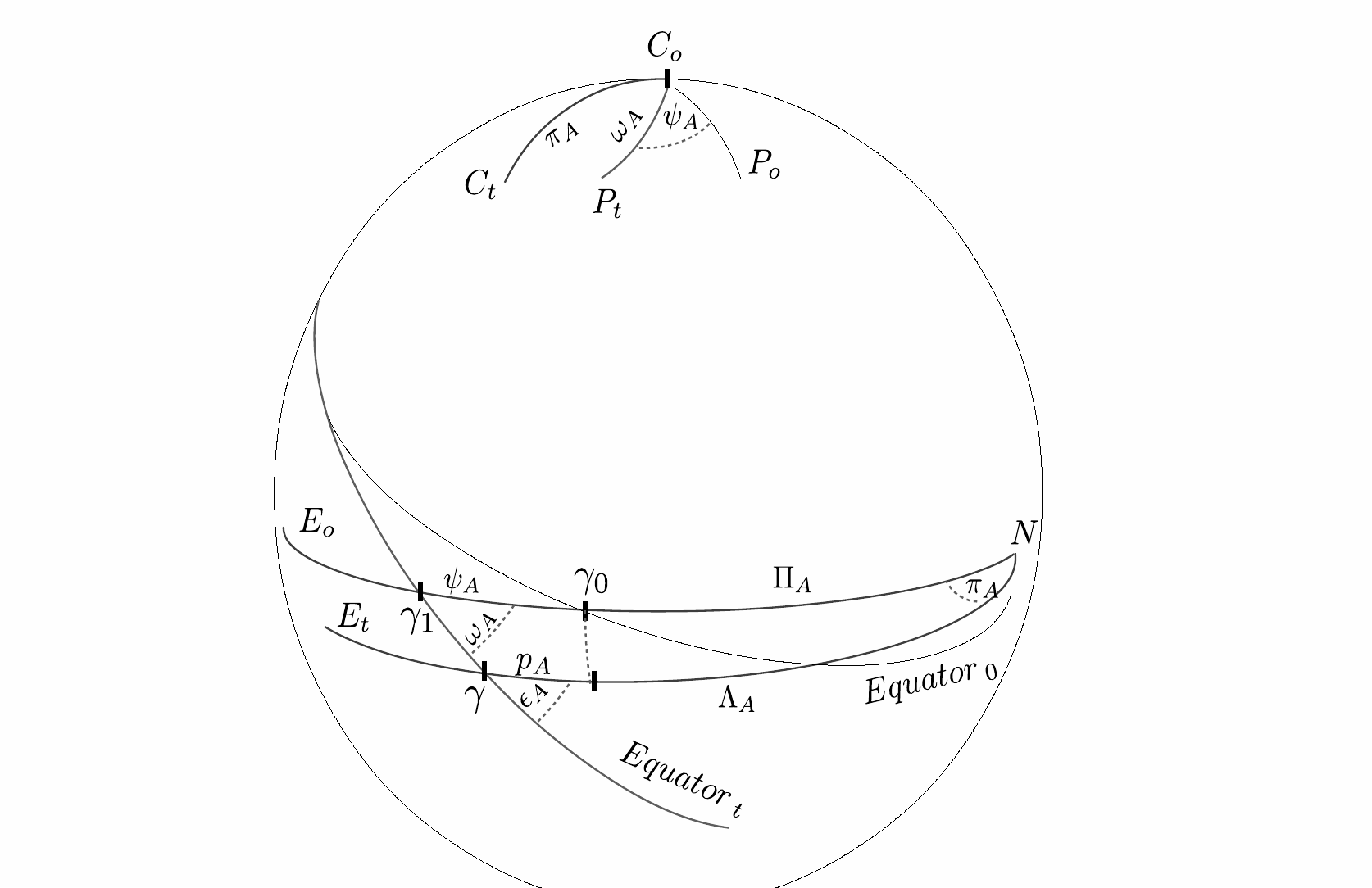}
\caption{Precessional elements measured with respect to the fiducial J2000.0 epoch. The mean equator of reference (J2000.0) and the mean equator 
of the date, are indicated. The nodal point $N$ is the intersection of the two ecliptics: $E_{{\rm o}}$ of the reference epoch, and $E_t$ the 
ecliptic of the date. The general precession in longitude accumulated from J2000.0 is $p_A$ (i.e., the segment from the projection of $\gamma_0$ 
on $E_t$, to $\gamma$ point). See the main text for descriptions of other angles.}
\end{center}
\end{figure}

\begin{figure}
\begin{center}
\includegraphics[scale=1]{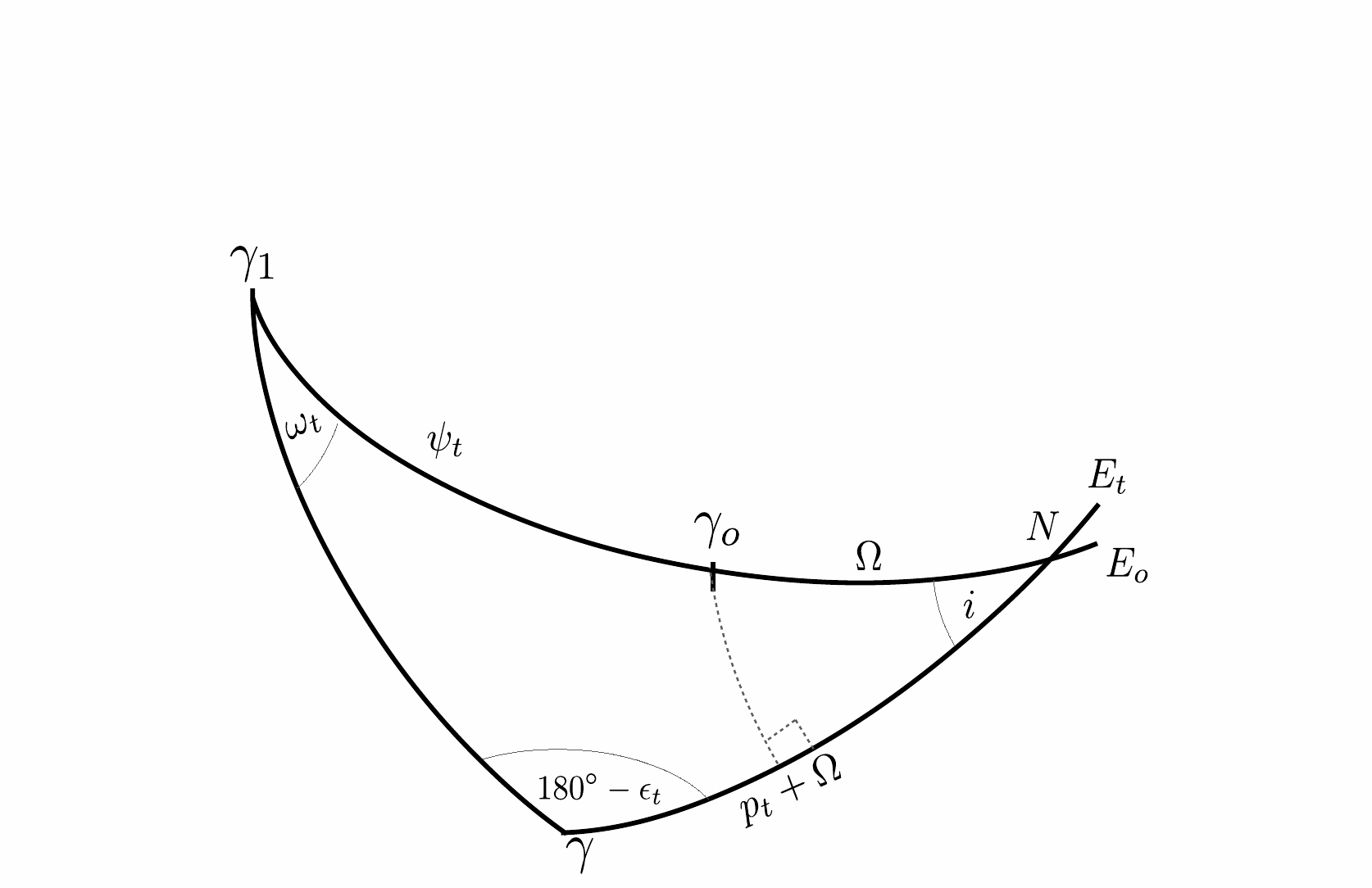}
\caption{ The spherical triangle used to find the ``true general precession'' and obliquity of the date is 
shown. The $\Omega$ and $i$ elements coming from DE431 define the plane of the Earth's orbit of the date with respect to the reference epoch 
(J2000.0). See the text for more details.}
\end{center}
\end{figure}

\newpage
\begin{figure}
\begin{center}
\includegraphics[scale=1]{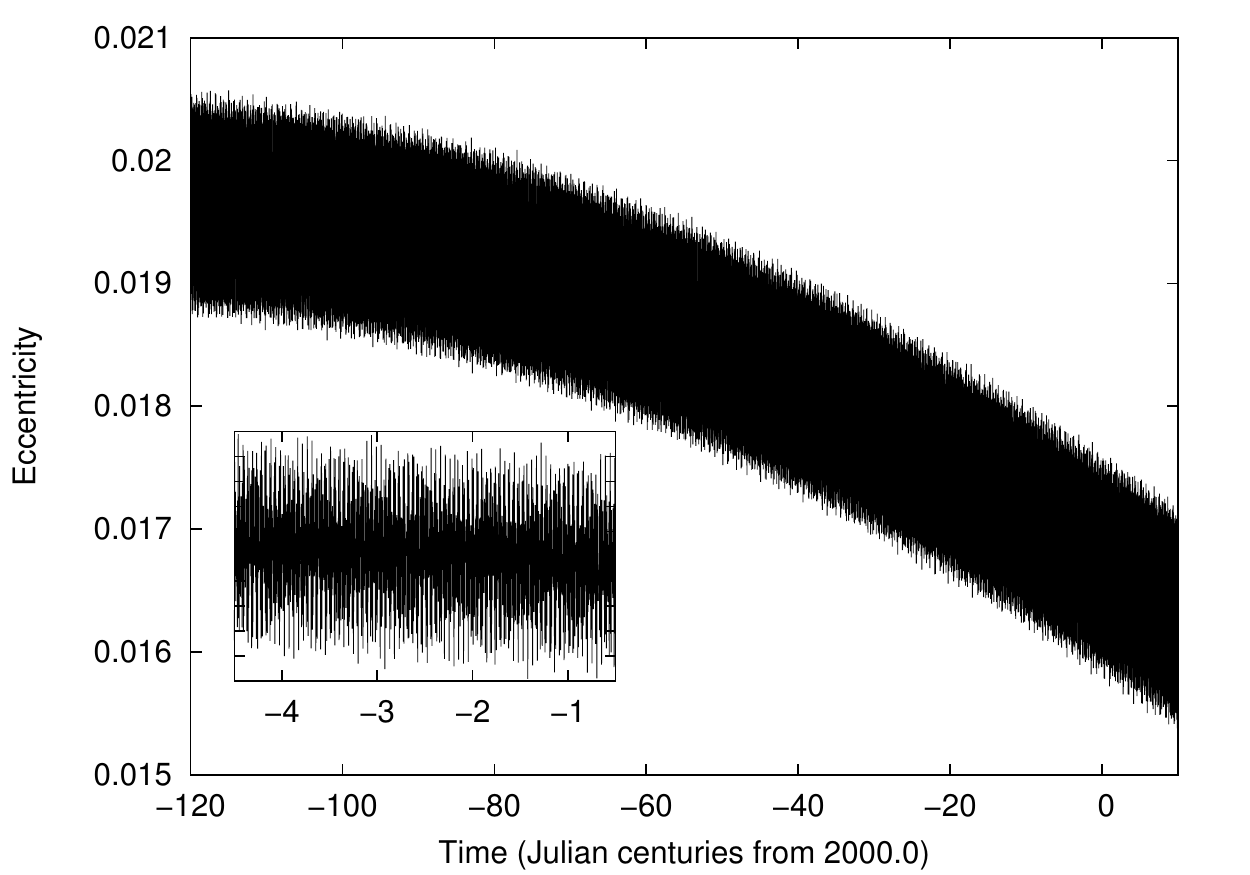}
\caption{Short-term evolution of the Earth's eccentricity over 13 kyr (one datum per 120 d). The last 500 yr, from 1950, is illustrated in the 
insert with expanded details.}
\end{center}
\end{figure}

\begin{figure}
\begin{center}
\includegraphics[scale=1]{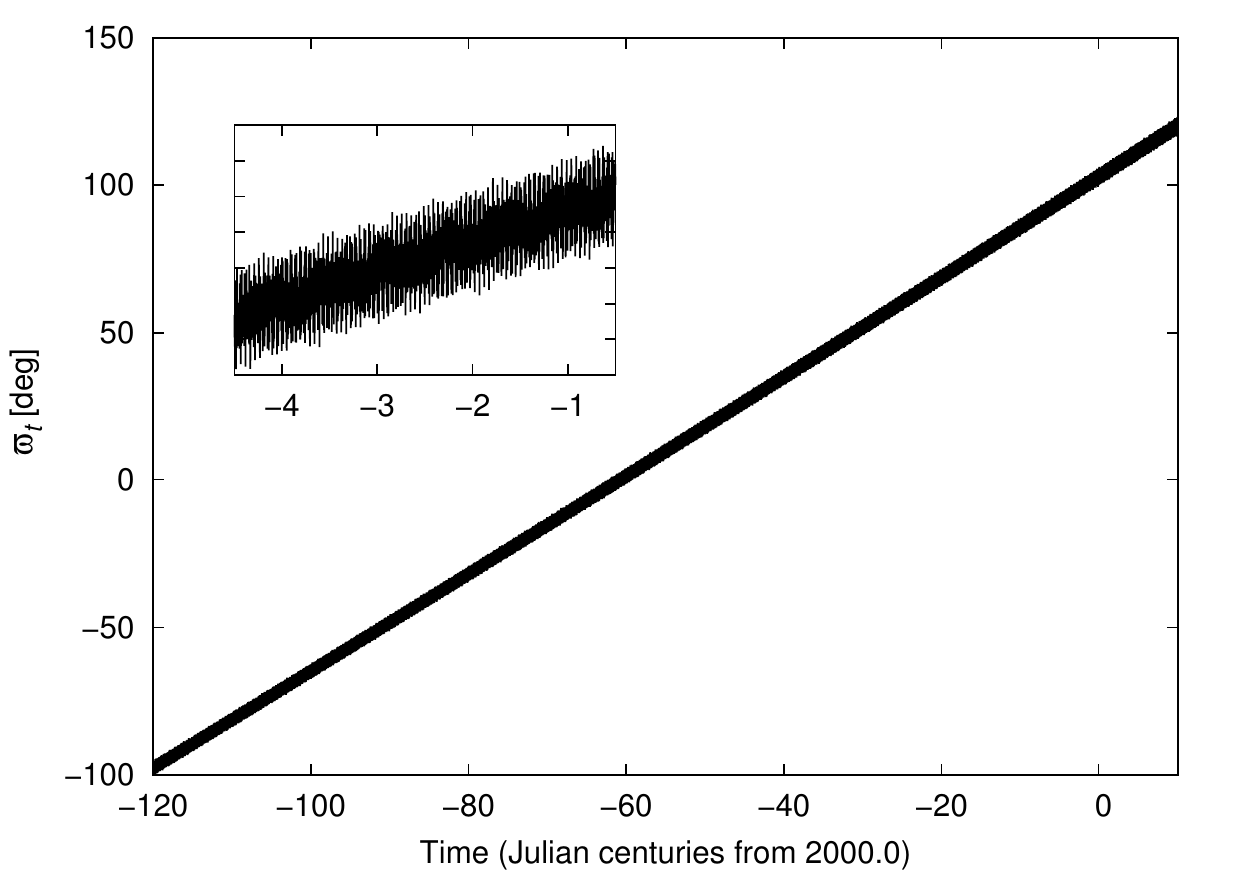}
\caption{Short-term evolution of the Earth's longitude of the perihelion over 13 kyr (one datum per 120 d). The last 500 yr, from 1950, is 
illustrated in the insert with expanded details.}
\end{center}
\end{figure}

\begin{figure}
\begin{center}
\includegraphics[scale=1]{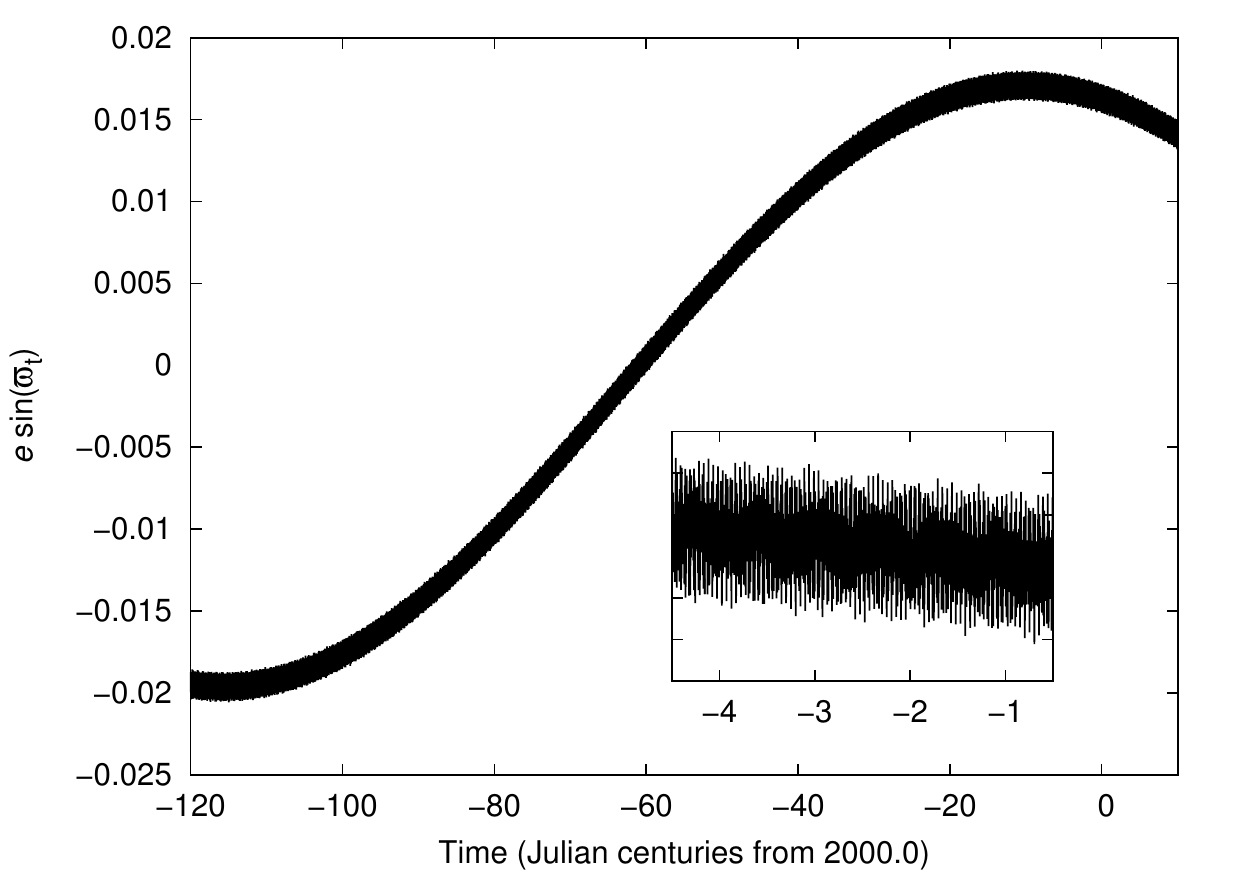}
\caption{Short-term evolution of the Earth's climatic precession  over 13 kyr (one datum per 120 d). The last 500 yr, from 1950, is illustrated 
in the insert with expanded details.}
\end{center}
\end{figure}

\begin{figure}
\begin{center}
\includegraphics[scale=1]{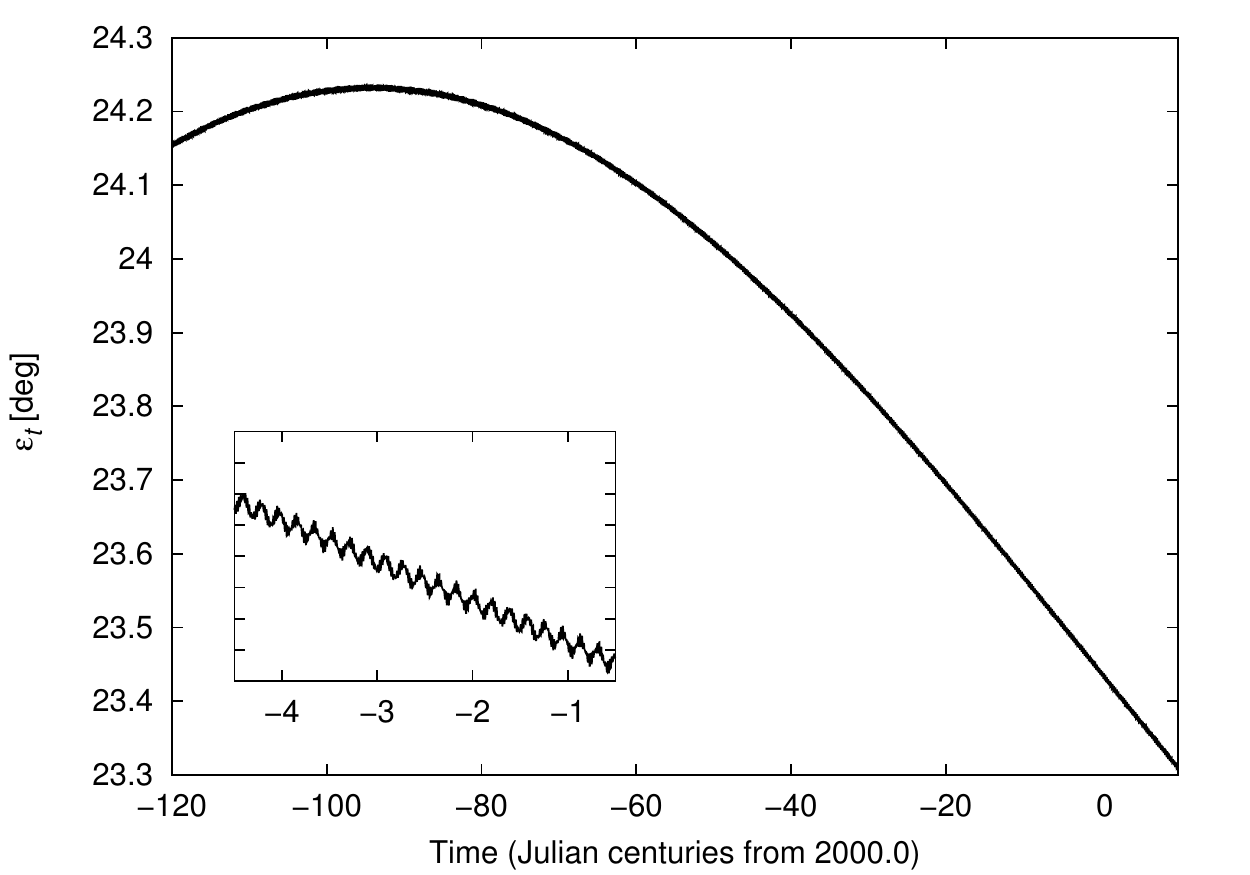}
\caption{Short-term evolution of the Earth's obliquity over 13 kyr (one datum per 120 d). The last 500 yr, from 1950, is illustrated in the 
insert with expanded details.}
\end{center}
\end{figure}

\begin{figure}
\begin{center}
\includegraphics[scale=1]{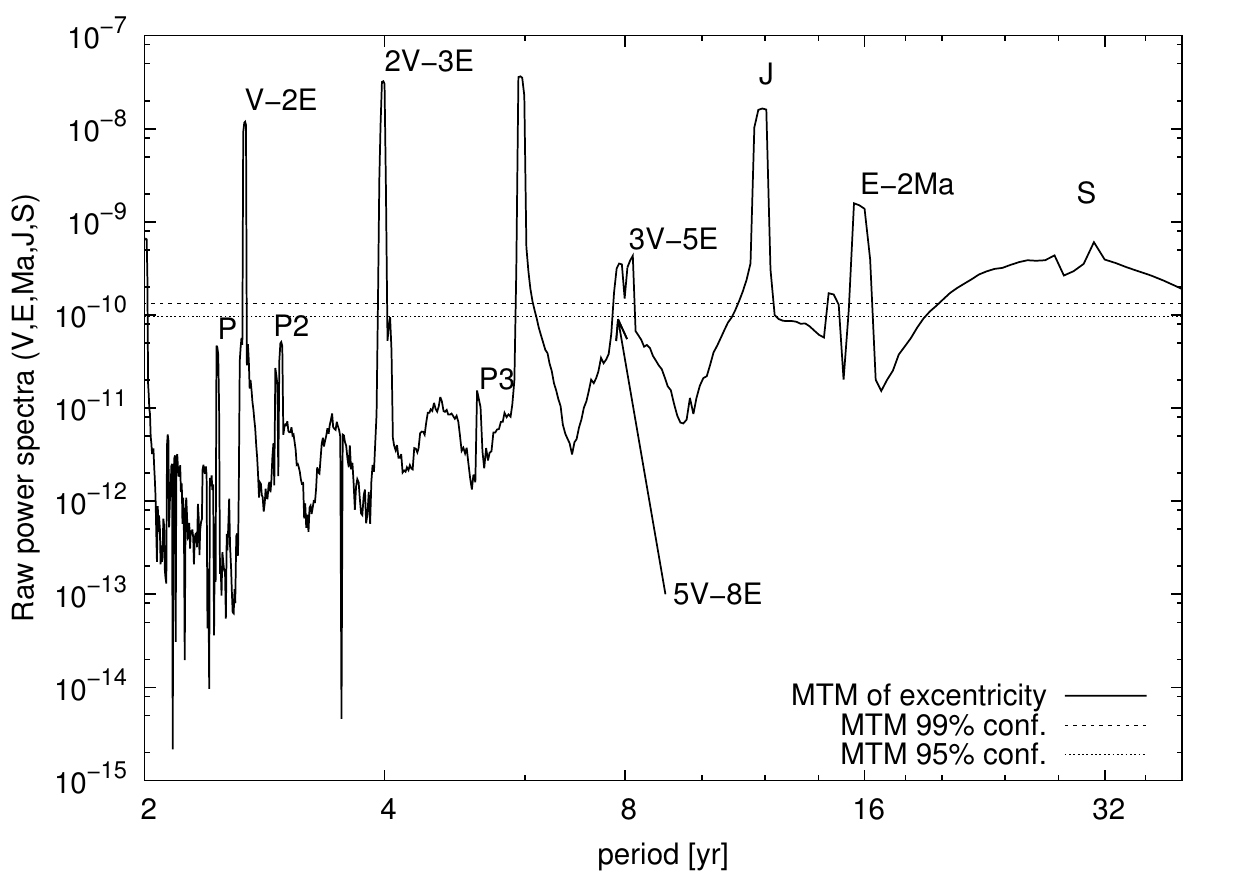}
\caption{MTM power spectrum of the raw data from the simulated perturbations on Earth's eccentricity by Venus to Saturn over 1 kyr (from 1000 to 
2000 AD). 
The labels mark the main recognized perturbations. P, P2 and P3 refer to perturbations peaks with confidence level less than 95\% 
which are related to planet Mars. 
One un-labeled peak between P2 and 2V-3E seems to be associated with the 4E-7Ma inequality.  
The other un-labeled but significant peaks are harmonics of labeled perturbations.}
\end{center}
\end{figure}

\begin{figure}
\begin{center}
\includegraphics[scale=1]{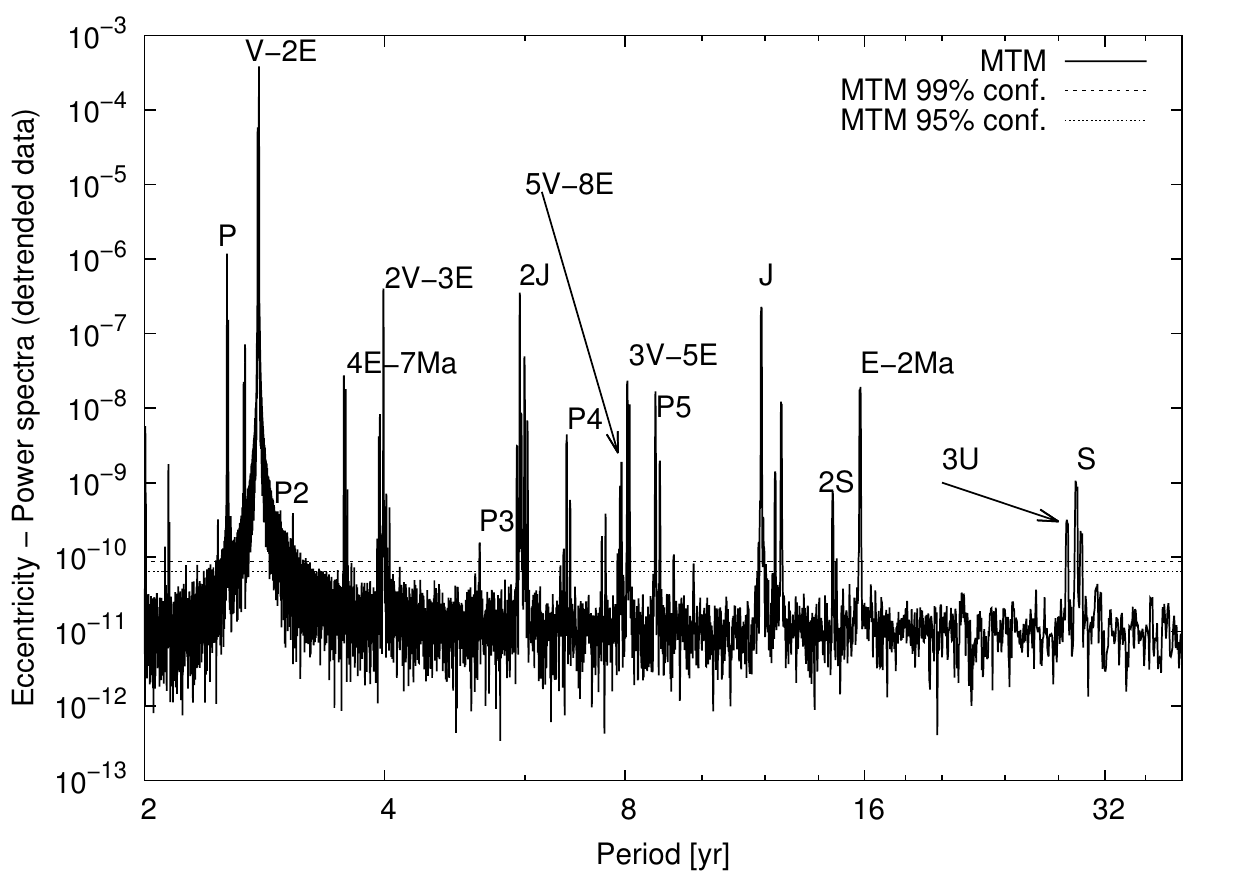}
\caption{MTM power spectrum for the short-term eccentricity variations based on the detrended data series after removing a parabolic long-term 
trend. The basic periodicities identified in Fig. 9 can also be detected here.}
\end{center}
\end{figure}

\begin{figure}
\begin{center}
\includegraphics[scale=1]{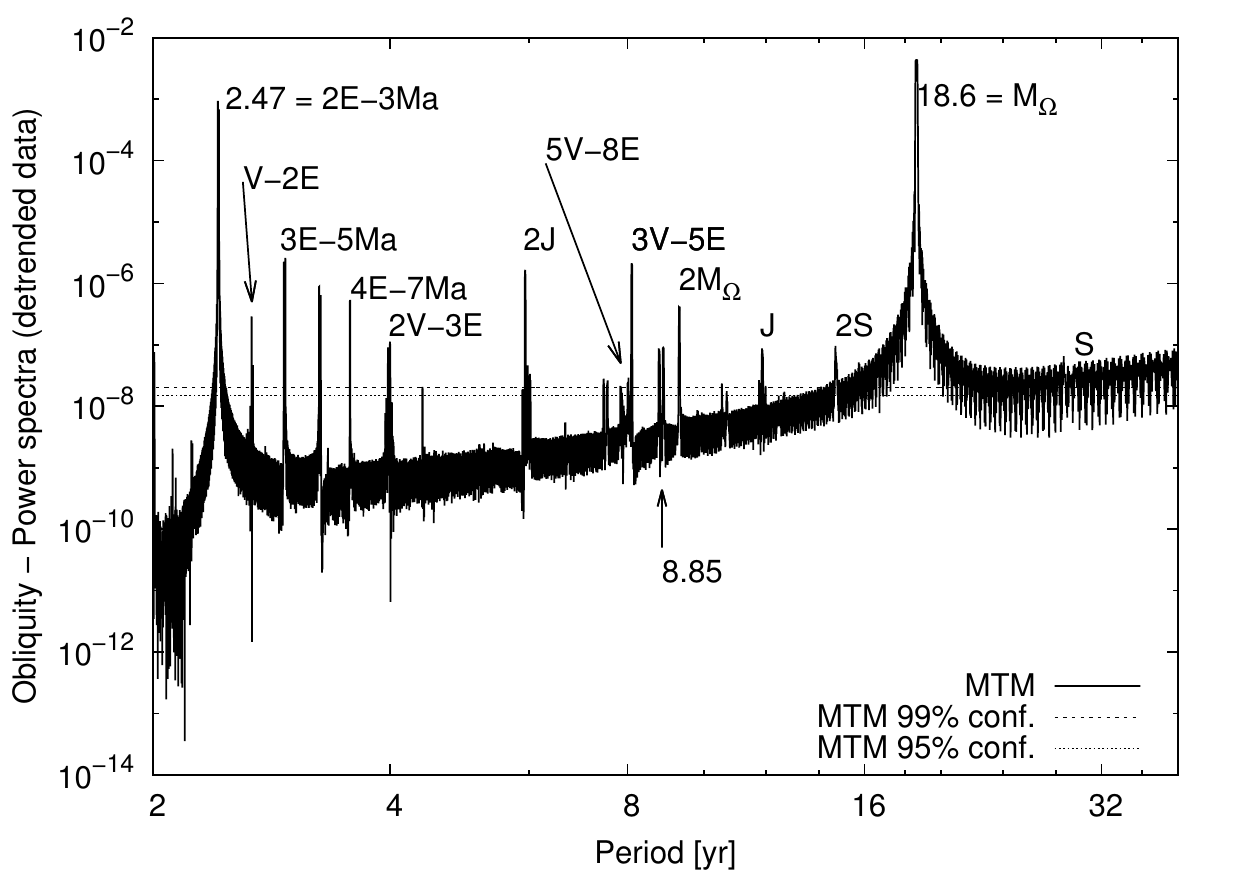}
\caption{ MTM power spectrum of the detrended obliquity time series. The main periodicities are identified (see discussion in the main text). 
Other periodicities are from harmonics and non-linear mixing of frequencies.}
\end{center}
\end{figure}

\begin{figure}
\begin{center}
\includegraphics[scale=1]{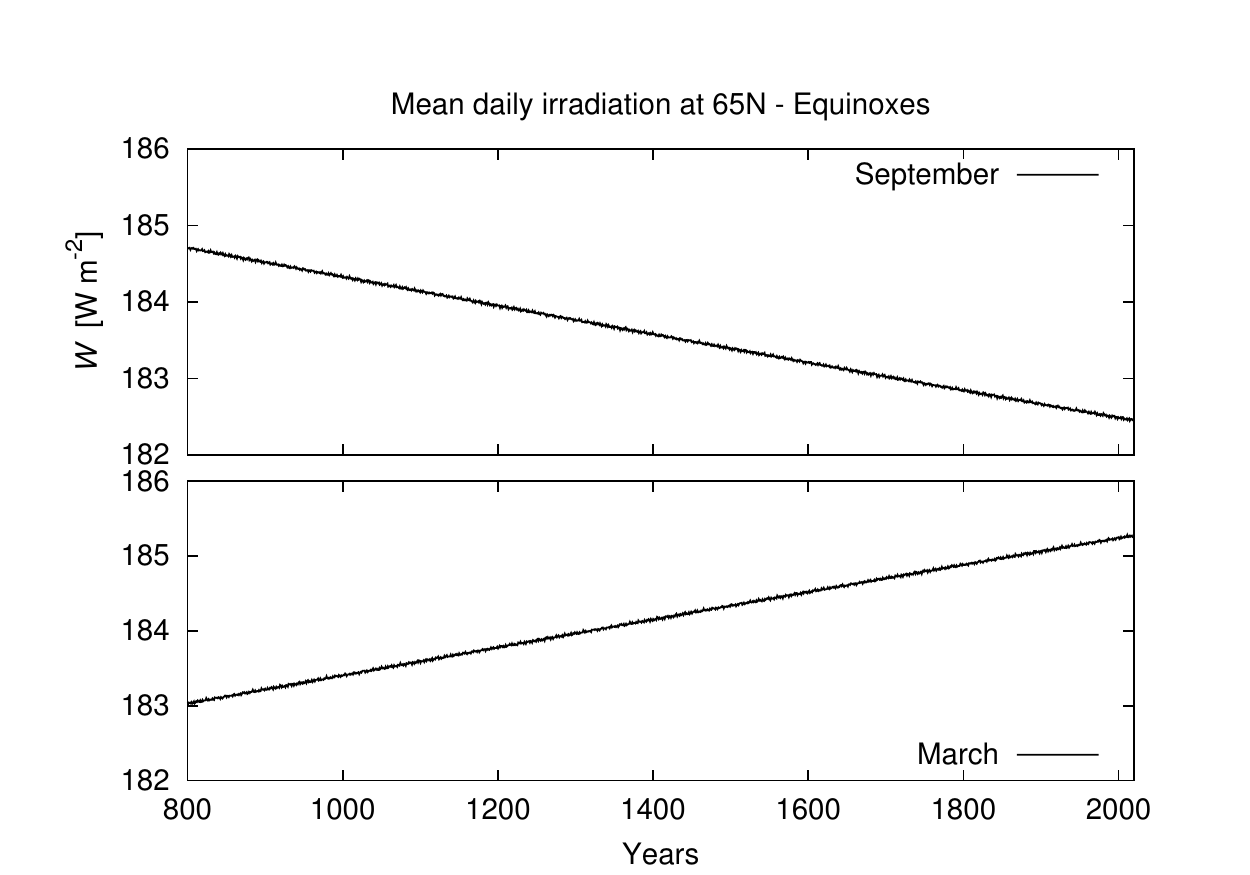}
\caption{Daily mean irradiation at $65^{\circ}$N over the indicated calendar interval for equinoxes; i.e., $\lambda_{\odot \, t} = 0,180$ deg. 
The solar constant used was 1366 W m$^{-2}$.}
\end{center}
\end{figure}

\begin{figure}
\begin{center}
\includegraphics[scale=1]{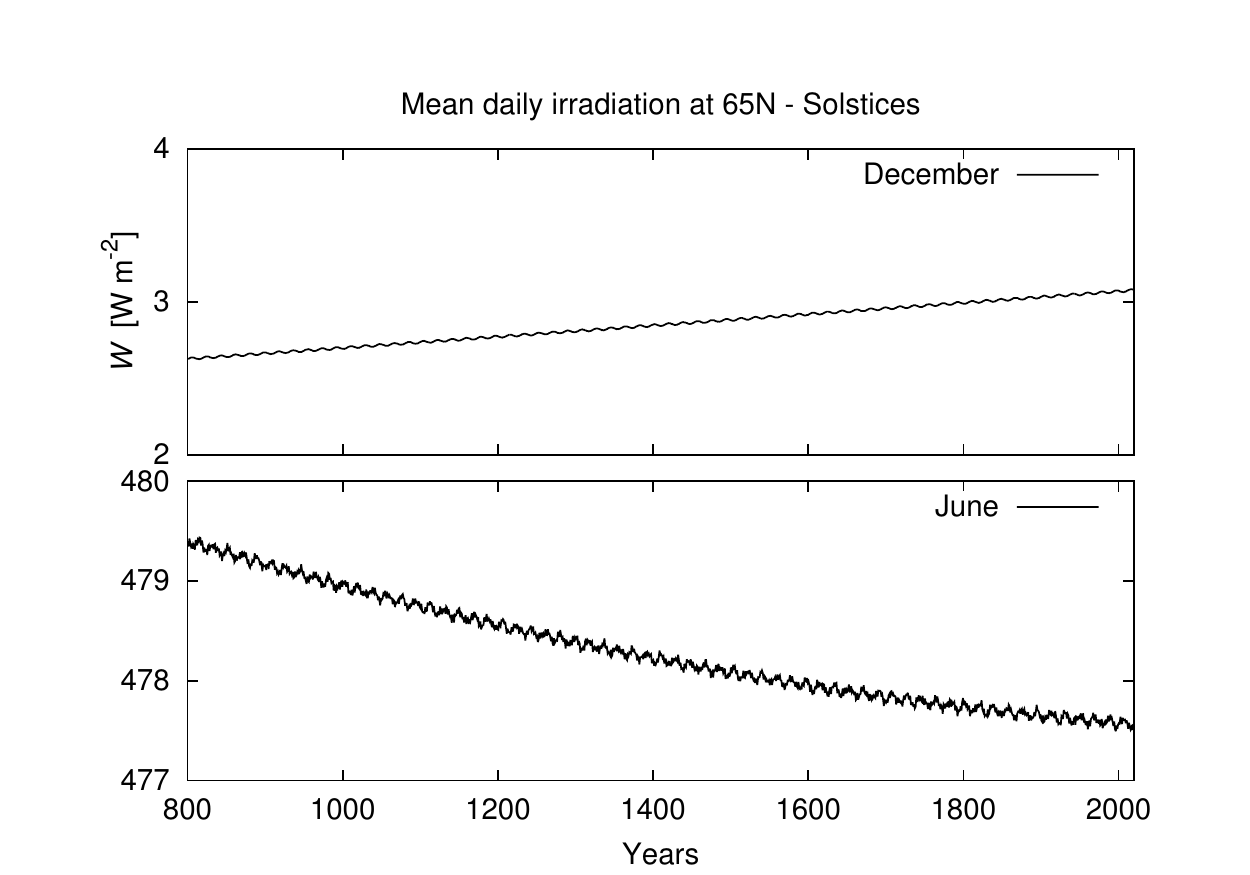}
\caption{The same as Fig. 12 but for solstices; i.e., $\lambda_{\odot \, t} = 90,270$ deg. The solar constant used was 1366 W m$^{-2}$.}
\end{center}
\end{figure}

\begin{figure}
\begin{center}
\includegraphics[scale=1]{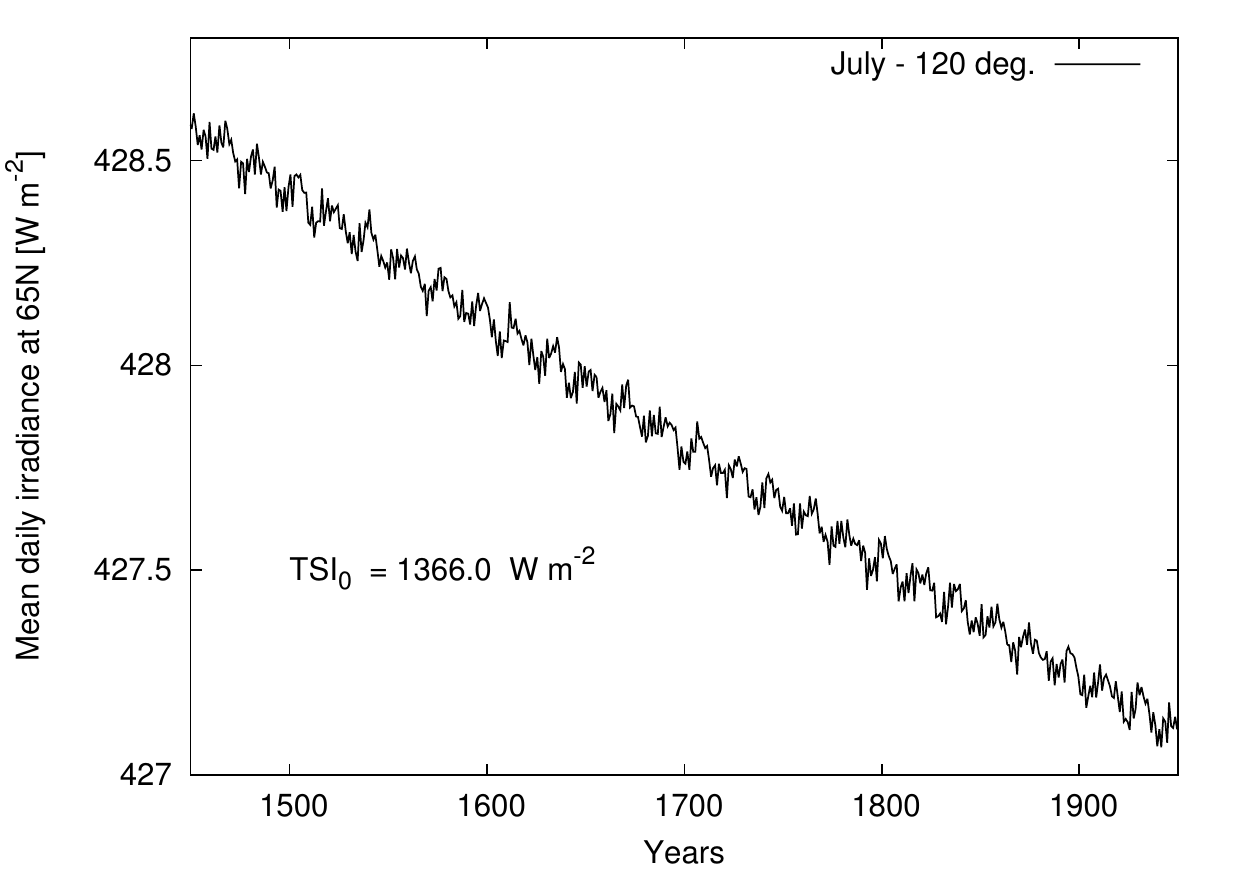}
\caption{$W$ at July mid-month; i.e., $\lambda_{\odot \, t}$ = 120 deg., for $65^{\circ}$N. The time interval corresponds closest to the one  
considered in \cite{Loutre1992}'s Fig. 12a. This comparison shows a very good agreement with their results. Solar constant used was 1366 W m$^
{-2}$.}
\end{center}
\end{figure}

\begin{figure}
\begin{center}
\includegraphics[scale=1]{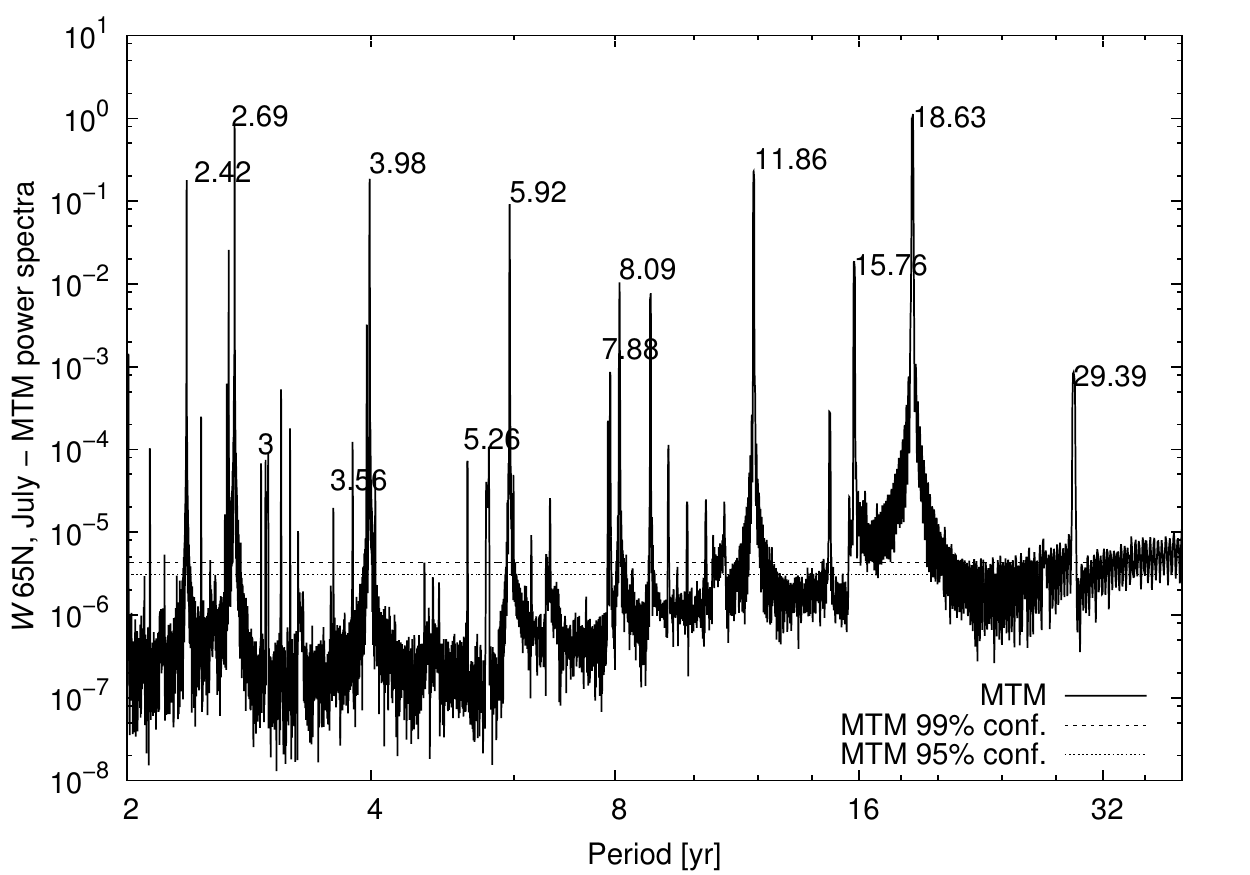}
\caption{ MTM spectrum of $W$ (over detrended time series) at July mid-month, $65^{\circ}$N. The main periodicities are identified and are 
described in the main text. At lower latitudes, the periodicities  arising from the obliquity signals are weakened.}
\end{center}
\end{figure}

\begin{figure}
\begin{center}
\includegraphics[scale=1]{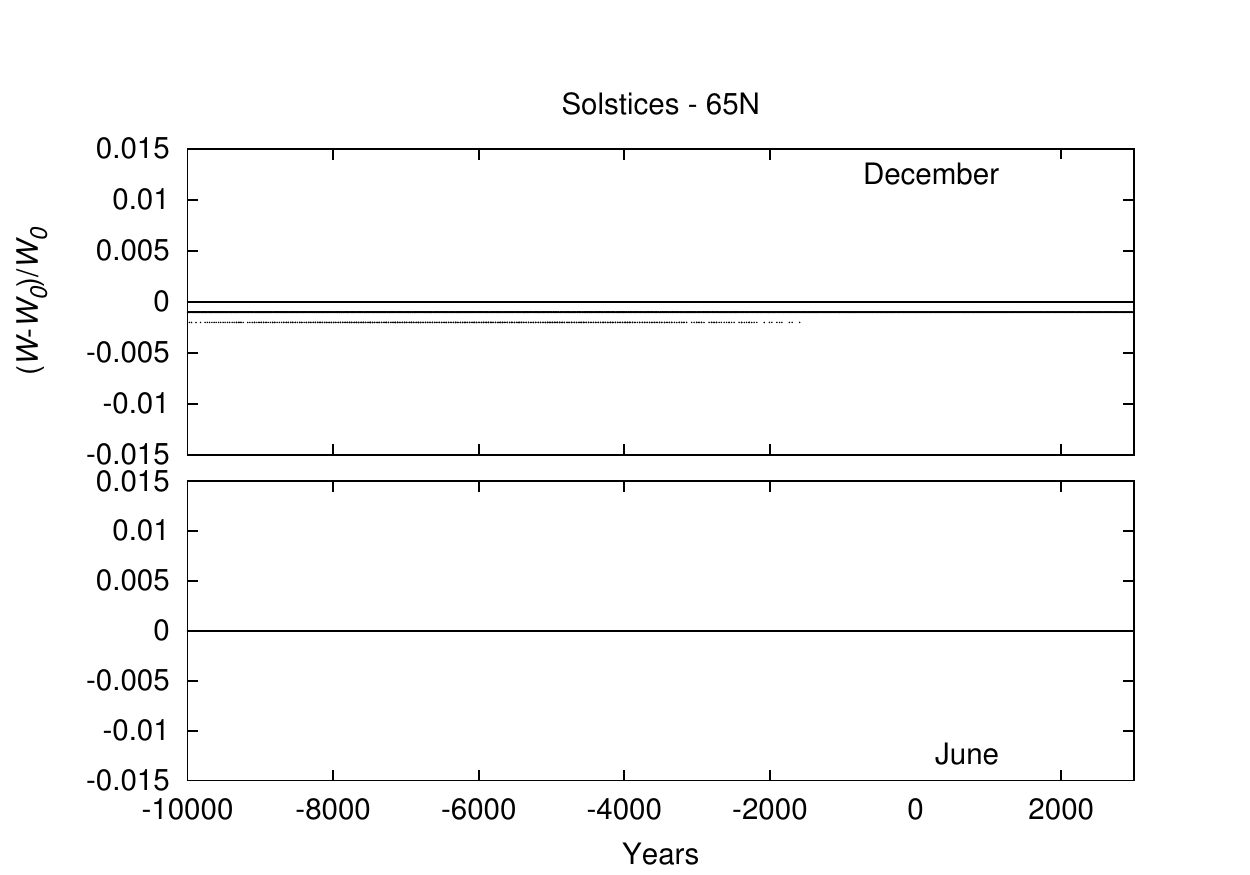}
\caption{Relative difference in $W$ calculation as a continuous $\lambda_{\odot \, t}$ or using tabulated values at mid-day, for solstices 
($\lambda_{\odot \, t}$ = 90, 270 deg.). 
At these moments, when the Sun is stationary, the errors are small, but 
they can reach 0.2\%. }
\end{center}
\end{figure}

\begin{figure}
\begin{center}
\includegraphics[scale=1]{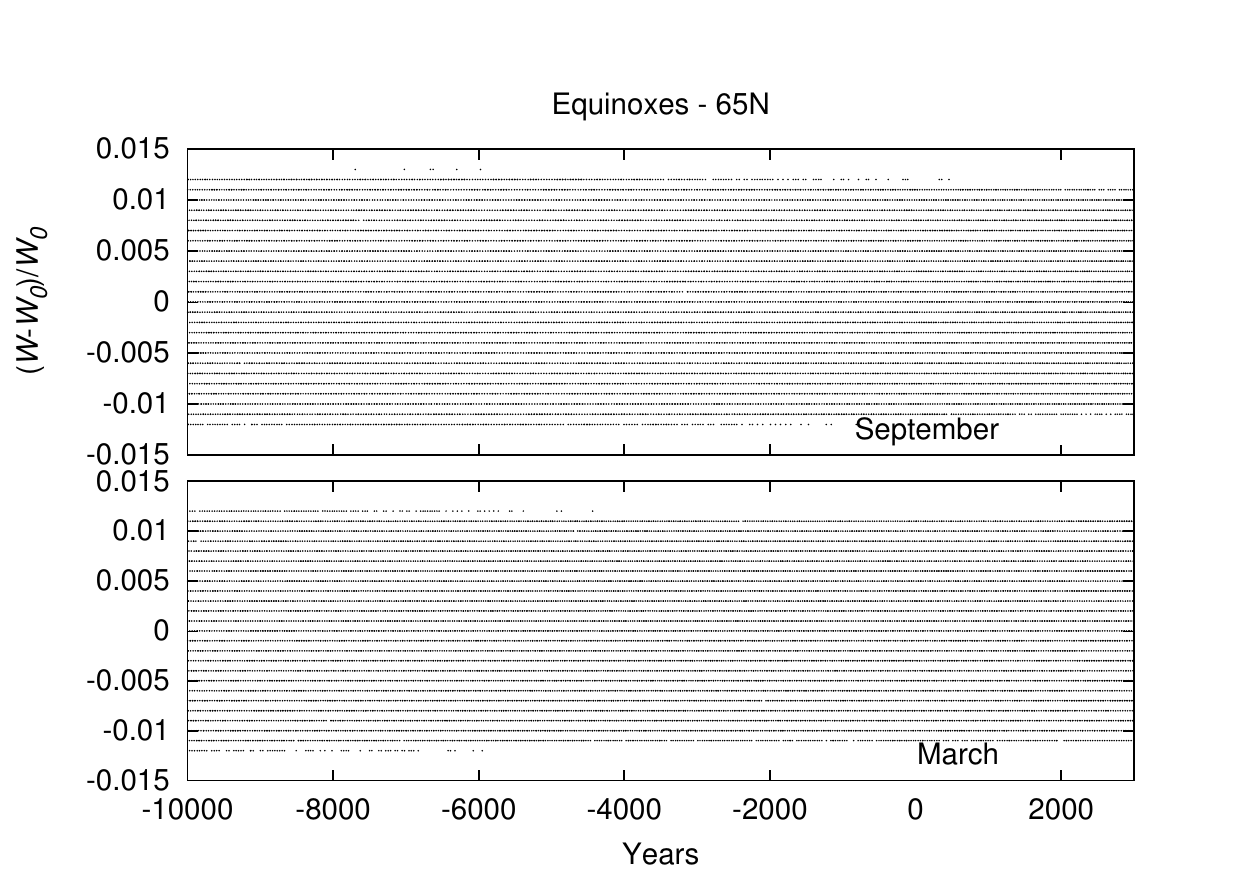}
\caption{The same as Fig. 16 but for  equinoxes ($\lambda_{\odot \, t}$ = 0, 180 deg.). The error is larger 
on average by one order of magnitude (or about 1.2 \%) when 
compared to the errors during solstices.}
\end{center}
\end{figure}

\clearpage
\begin{figure}
\begin{center}
\includegraphics[scale=1]{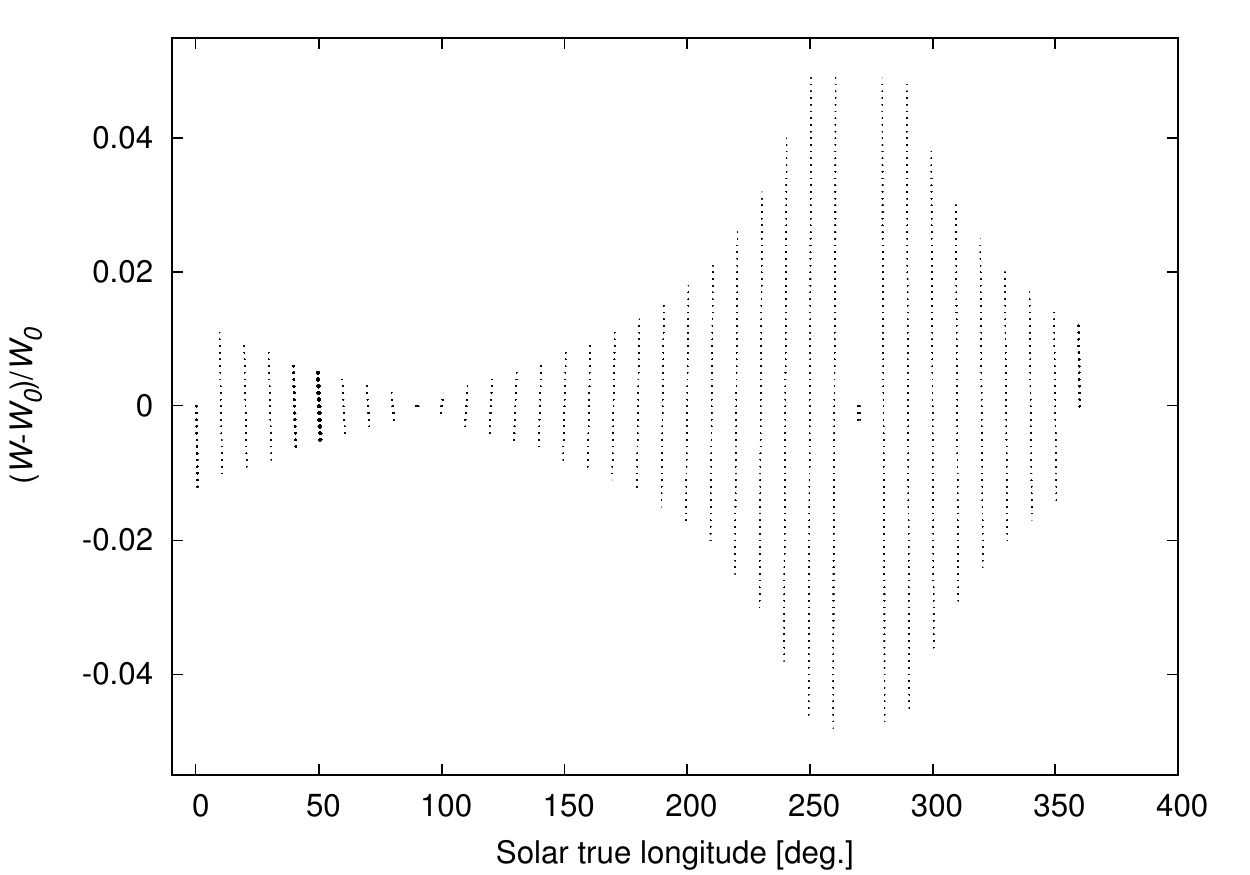}
\caption{Relative difference in the $W$ calculation as a function of $\lambda_{\odot \, t}$ over the 13 kyr interval we studied (for 65$^{\circ}
$N). 
Around the solstice of December, the relative error can reach 5\% but drop to the level of less than 0.1\% at 270 deg.} 
\end{center}
\end{figure}

\begin{figure}
\begin{center}
\includegraphics[scale=1]{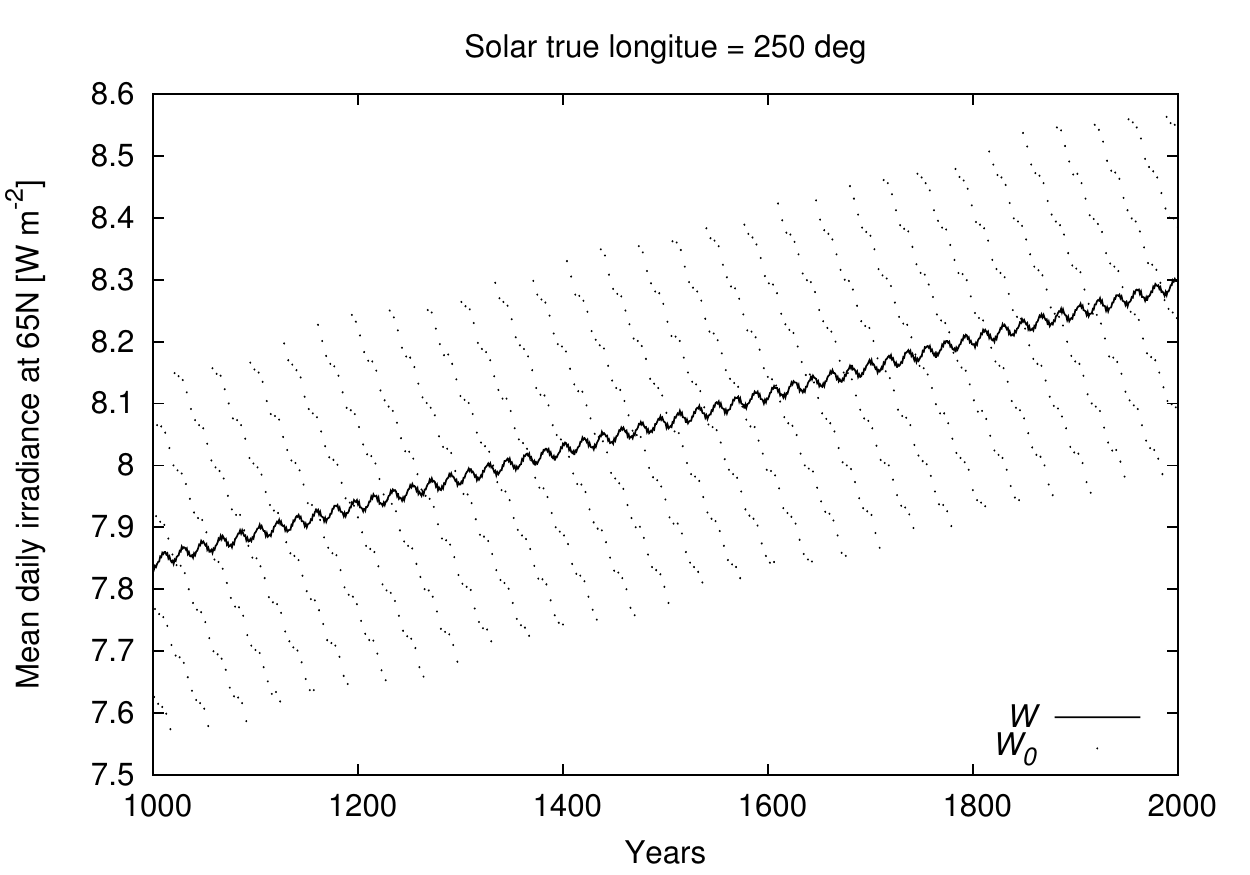}
\caption{$W$ calculation as a continuous $\lambda_{\odot \, t}$ (lines) or using tabulated values at mid-day (dots), for $\lambda_{\odot \, t}$= 
250 deg. (for 65$^{\circ}$N),  from 
1000 to 2000 AD. The absolute difference is less than 0.3 W m$^{-2}$.}
\end{center}
\end{figure}

\begin{figure}
\begin{center}
\includegraphics[scale=1]{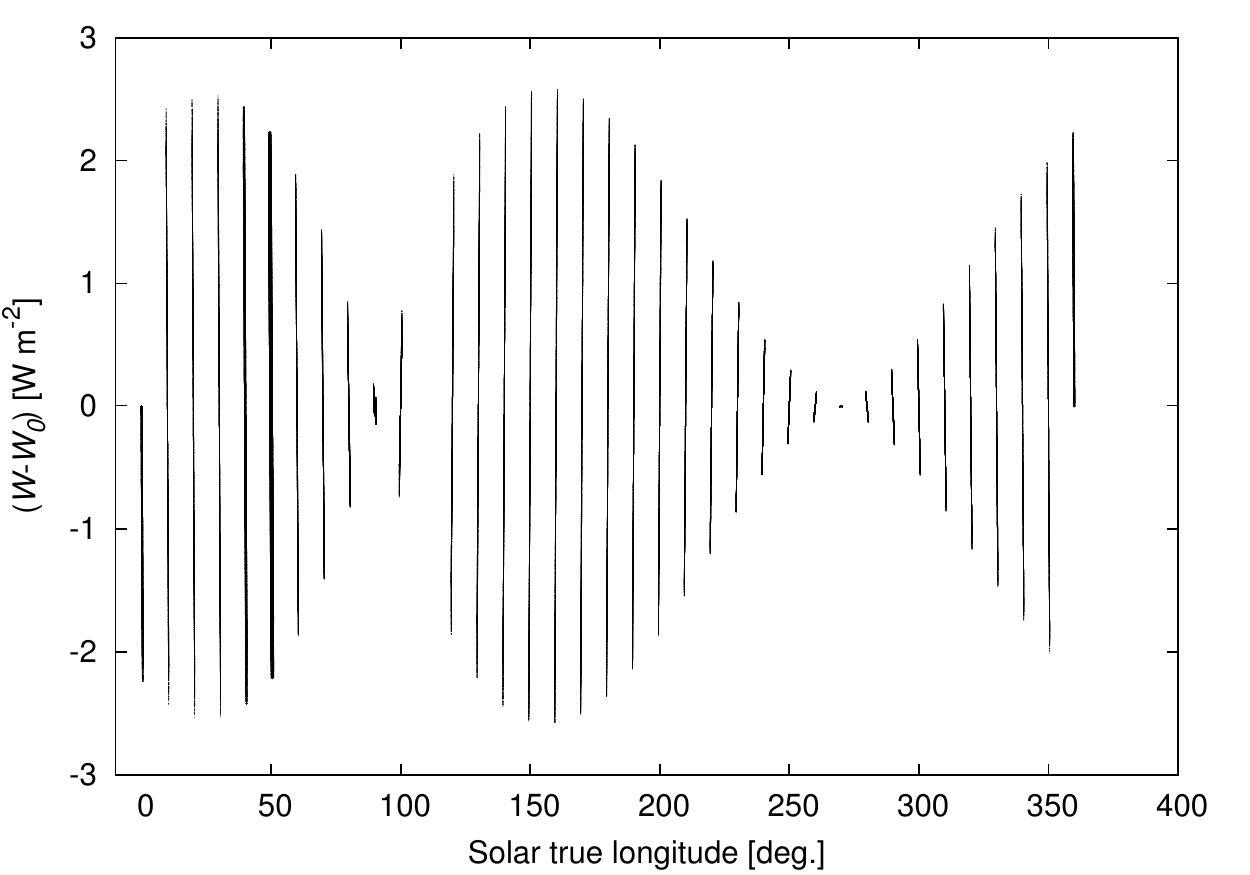}
\caption{Difference $W-W_0$ in mean daily insolation calculation
 as a function of $\lambda_{\odot \, t}$ for the 13 kyr interval we studied (for 65$^{\circ}$N).
At boreal spring and at the end of the boreal summer, the absolute difference can
reach 2.5 W m$^{-2}$ (see the text for more explanation). }
\end{center}
\end{figure}

\begin{figure}
\begin{center}
\includegraphics[scale=1]{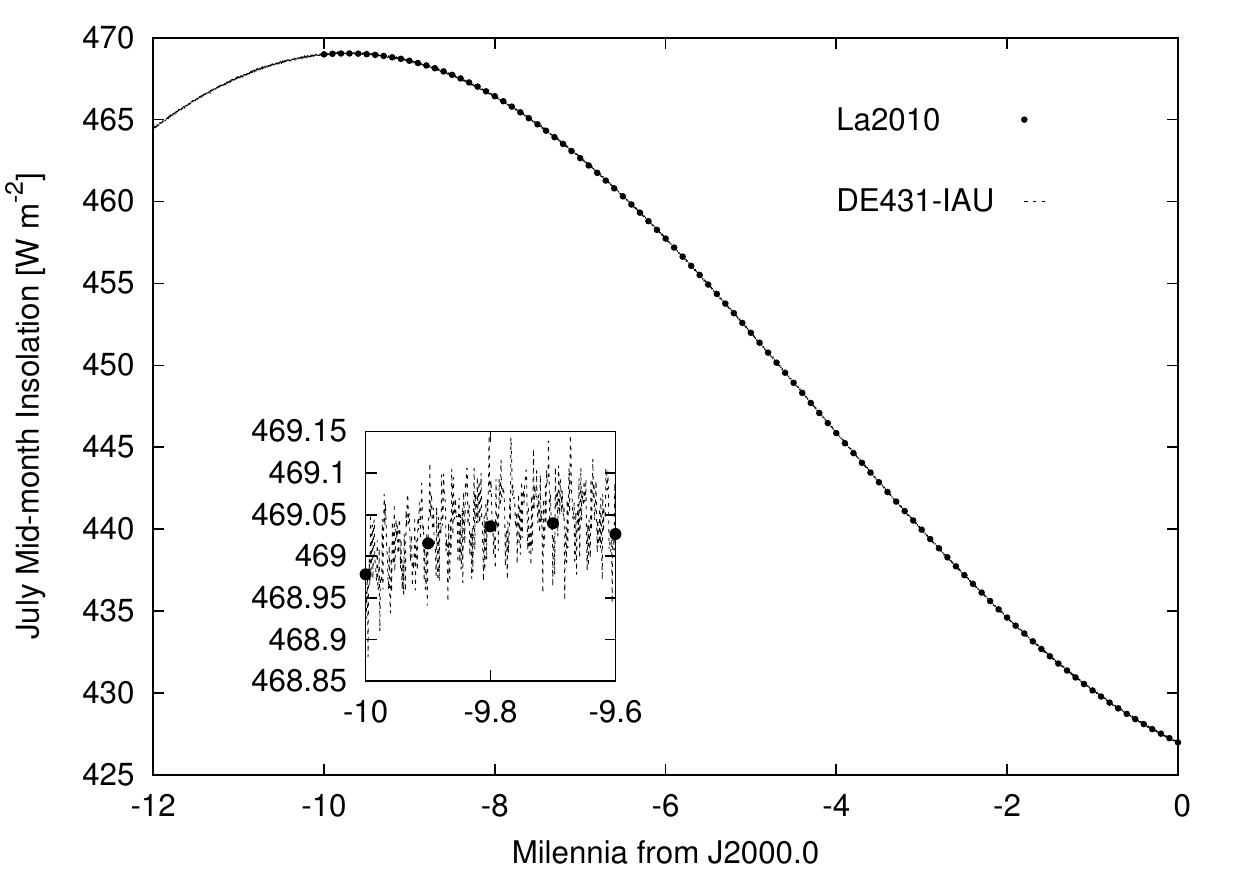}
\caption{$W$ at July mid-month ($\lambda_{\odot \, t}$ = 120 deg.), for 65$^{\circ}$N: this work, DE431-IAU (dashed lines); Laskar et al. (2011a) 
solution, La2010 (points). The comparison is shown  starting at $-$10 kyr with expanded details of the intercomparison illustrated in the insert.
The solar constant used was 1366.0 W m$^{-2}$.}
\end{center}
\end{figure}

\begin{figure}
\begin{center}
\includegraphics[scale=1]{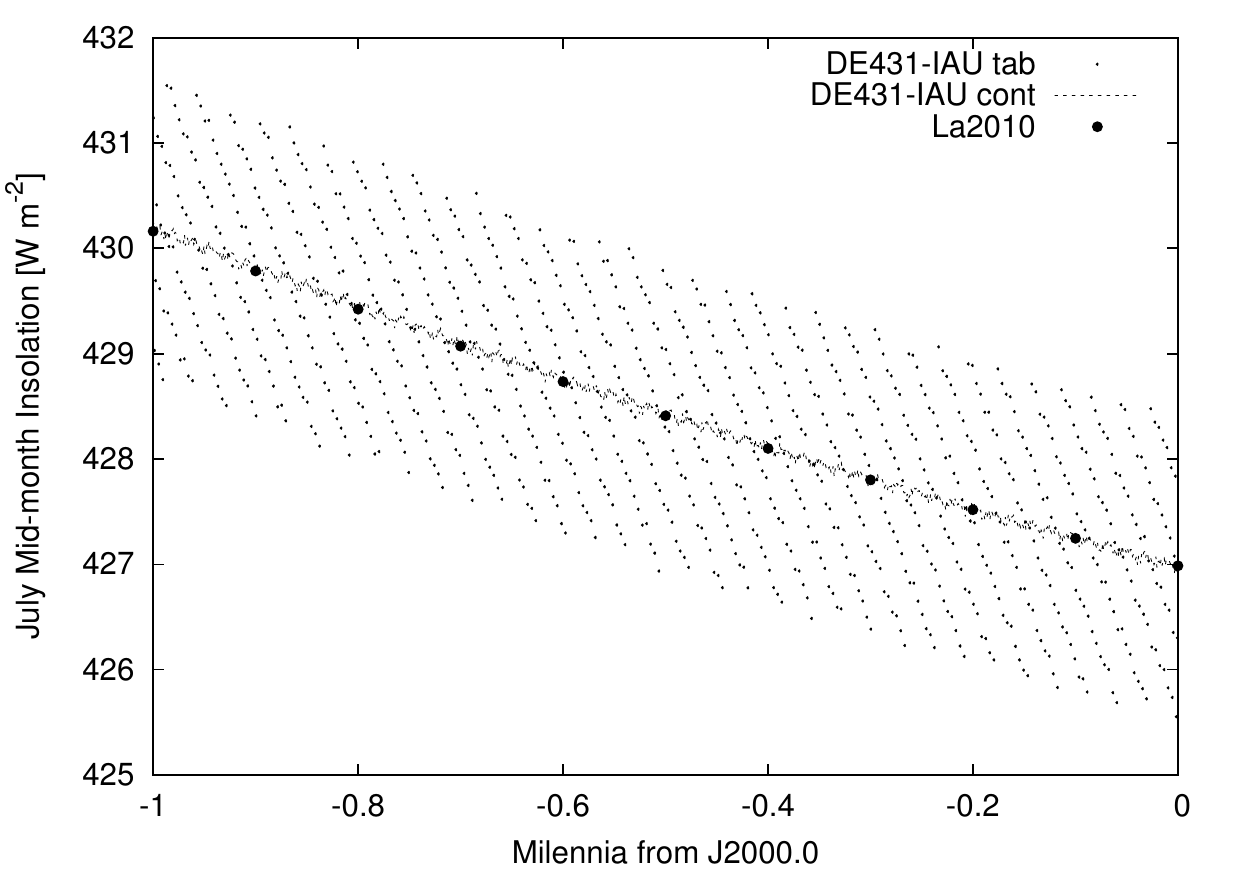}
\caption{$W$ at July mid-month ($\lambda_{\odot \, t}$ = 120 deg.), for 65 $^{\circ}$N: DE431-IAU solution, continuous values (dashed lines); 
La2010 data (points) and DE431-IAU solution, with tabulated values (dots). Differences between the continuous and 
tabulated values can reach $\pm$ 0.3\%; i.e., approximately $\pm$ 1.2 W m$^{-2}$.}
\end{center}
\end{figure}

\end{document}